\documentclass[preprint,12pt,authoryear]{elsarticle}

\usepackage{amssymb}
\usepackage{graphicx}       
\usepackage{subfigure}
\usepackage{amsthm}
\usepackage{amsmath}
\usepackage{lineno}
\usepackage{array}
\usepackage{booktabs}
\usepackage{multirow}
\usepackage{tabularx}
\usepackage{bm}
\usepackage{float}
\usepackage{color}
\allowdisplaybreaks[4]
\UseRawInputEncoding
\graphicspath{{figures/}}
\usepackage[dvipsnames]{xcolor}
\usepackage[colorlinks=true,
linkcolor=Maroon,
urlcolor=blue,
citecolor=blue]{hyperref}

\journal{}

\begin{document}

\begin{frontmatter}



\title{Particle resolved simulation of sediment transport by a hybrid parallel approach}


\author[xidian,minnesota]{Zhengping Zhu}
\author[lzu]{Ruifeng Hu}
\author[lzu]{Yinghaonan Lei}
\author[minnesota]{Lian Shen}
\author[xidian,lzu]{Xiaojing Zheng\corref{cor1}}
\ead{xjzheng@xidian.edu.cn}
\cortext[cor1]{Corresponding author}

\address[xidian]{Research Center for Applied Mechanics, School of Mechano-Electronic Engineering, Xidian University, Xi'an 710071, China}
\address[lzu]{Center for Particle-Laden Turbulence, Key Laboratory of Mechanics on Disaster and Environment in Western China, Ministry of Education, and College of Civil Engineering and Mechanics, Lanzhou University, Lanzhou 730000, China}
\address[minnesota]{St. Anthony Falls Laboratory and Department of Mechanical Engineering, University of Minnesota, Minneapolis, Minnesota 55455, USA}

\begin{abstract}
{Sediment transport over an erodible sediment bed is studied by particle resolved simulations with a hybrid parallel approach. To overcome the challenges of load imbalance in the traditional domain decomposition method when encountering highly uneven distributions of particles in sediment transport, the parallel approach of \cite{darmana2006parallelization} originally developed for point particle simulations is modified and implemented into particle resolved simulations.} A novel memory optimization technique is proposed to reduce the memory requirement of the hybrid approach for spherical particles with equal size. The present hybrid parallel approach shows good scalability and high parallel efficiency in a challenging sediment transport test case with more than a million spherical particles. Our code is validated by several benchmark cases, and the results show good agreement with experimental and computational data in the literature. Furthermore, a turbulent flow over an erodible sediment bed is simulated. An extraction method is proposed to distinguish the saltating and rolling particles and extract impact and rebound information of the particle-mobile bed interaction. The probability distribution functions (PDF) of several saltation parameters such as velocity, angle, and spanwise angular velocity of impact and rebound events are presented. Splash functions are established for the particle-mobile bed interaction in the turbulent flow, which was rarely investigated in the experiments and is helpful to model the complex particle-bed interactions in turbulent flow.
\end{abstract}


\end{frontmatter}




\newpage

\section{Introduction}
When an erodible sediment bed is exposed to a sufficiently strong shear flow, bed particles may be entrained and transported by the flow under the combined action of the hydrodynamic force, gravitational force, and interparticle contact force, a process known as sediment transport \cite[]{bagnold1941physics,graf1984hydraulics,chien1999mechanics,shao2008physics,zheng2009mechanics}. Sediment transport by wind or water is ubiquitous in nature, such as in sand/dust storms, debris flows, rivers, and coastal environments. It is one of the most important geophysical processes responsible for wind erosion, dust aerosol emission, and the formation of dunes and ripples. A deep understanding of the mechanism of particle-fluid interactions over erodible sediment beds is vital for accurate prediction of sediment transport and geomorphological variations.

\textcolor{black}{There have been extensive experimental and numerical studies on sediment transport over erodible sediment bed \cite[]{merritt2003review,le2007sediment,papanicolaou2008sediment,duran2011aeolian,kok2012physics,valance2015physics,yang2017,zhang2018large,zhu2019sand,pahtz2020physics,ranaentrainment}. The interactions between the particle bed and turbulence in sediment transport have always been the focus of studies \cite[]{ho2011scaling,lanigan2016atmospheric,bo2017improved,berk2020transport,bragg2021mechanisms,zheng2021modulation,zheng2021experimental,zhu2021AE}. In the experiments, it is hard to measure particle-mobile bed interaction (i.e., impact and rebound information) directly since the impact particles near the bed are hard to be distinguished from the moving particles in the fluid and the mobile sediment bed. For this reason, the stochastic features of particle-mobile bed interaction were investigated by the inversion from the bouncing particle trajectories in experiment \cite[]{nino1994gravel,nino1998using,lee2006three} or by an incident particle colliding with a static bed \cite[]{mitha1986grain,werner1990steady, tanaka2002discrete,ammi2009three,chen2019experimental}. Although the interaction between particle and mobile sediment bed can be directly obtained in numerical simulation by continuously recording all particles' motion, the effect of turbulence was usually not considered in the most of the simulations \cite[]{anderson1988simulation,anderson1991wind,shao1999numerical,huang2003effects,kok2009comprehensive,duran2012numerical,berzi2016periodic}, which may affect the particle-mobile interaction. The complex physics of the particle-mobile bed interaction in the turbulent flow of sediment transport calls for high-fidelity simulation study. Among different numerical simulation methods, including the Eulerian method, Lagrangian point-particle method, and particle resolved method, the particle resolved direct numerical simulation has the highest fidelity \cite[]{uhlmann2005immersed,luo2007full,yu2007direct,breugem2012second,kempe2012improved,tenneti2014particle,zhou2014second,picano2015turbulent,akiki2016immersed,tschisgale2017non,wang2017modulation,costa2018effects,tao2018combined,peng2019direct,wang2019hydrodynamic}. Benefiting from the advancements in high-performance computation in recent years, particle resolved direct numerical simulations (PRDNS) of sediment transport have become feasible. \cite{ji2014saltation} investigated the statistical features of saltation particles by the PRDNS of sediment transport. \cite{kidanemariam2014direct,kidanemariam2014interface,kidanemariam2017formation} investigated the formation of sediment patterns in the sediment transport by PRDNS. \cite{vowinckel2016entrainment} investigated the mechanism of particle entrianment over an erodible bed by the PRDNS of sediment transport. \cite{jain2021impact} investigated the sediment transport with different shape of particles by the PRDNS. However, these particle resolved simulations did not focus on the particle-mobile bed interaction in the turbulent flow of sediment transport. 
}

\textcolor{black}{The particle-resolved simulation is also the most computationally intensive method because it needs to fully resolve the scales of turbulence and particles simultaneously. Moreover, in sediment transport, particle collisions need to be modeled, which significantly increases the computational complexity. The high computational cost and memory consumption of the PRDNS pose great challenges to its utilization in basic and applied research. Further improvements of the parallel algorithms and memory optimization techniques are critically needed. Most of the previous parallel algorithms are developed for the point-particle simulations \cite[]{uhlmann2004simulation,darmana2006parallelization,tsuji2008spontaneous,kafui2011parallelization,gopalakrishnan2013development,amritkar2014efficient,wang2017parallel,pozzetti2019parallel,dufresne2020massively}. These methods can not be directly applied to particle resolved simulations since the particle is treated as a finite size rather than a sizeless point in the particle resolved simulation. It causes differences in the treatment of particle-fluid interactions. In the particle resolved simulation, the surface of a finite-sized particle is represented by Lagrangian points, and particle-fluid interactions are usually calculated by the immersed boundary (IB) method or the fictitious domain method \cite[]{uhlmann2005immersed,yu2007direct,luo2007full,kempe2012improved,breugem2012second}. The feedback force of a particle on the fluid is determined by interpolating the forces on the Lagrangian points to the Eulerian grid cells. The force can also affect other Eulerian grid cells surrounding the one where the Lagrangian point is located.
If the domain decomposition method is employed, then the force may act on a ghost cell of another subdomain when a Lagrangian point is near the subdomain boundary. In this case, the force in the ghost cell needs to be mirrored back to the adjacent subdomain, which increases the complexity of the parallelization of the particle resolved simulation compared with the point-particle simulation.}

\textcolor{black}{Currently, only a few parallel methods are developed for the particle resolved simulations. \cite{uhlmann2004simulation} proposed a master-slave parallel approach for the particle resolved simulations. Both the carrier and disperse phase are parallelized with the domain decomposition. The master processor is defined as the processor where the particle center resides. The surrounding processors occupied by the finite-size particles are defined as salve processor. The data transmit between master and slave processors is complex. \cite{wang2013parallel} proposed a 'gathering and scattering' strategy parallel approach for the particle resolved simulations. \cite{valero2014accelerating} proposed a parallel approach on multi-core and GPU architectures for the particle resolved simulations. \cite{yu2006fictitious} proposed a parallel fictitious domain method for particle resolved simulations. \cite{yang2021scalable} proposed a highly scalable parallel approach named DBGP for the particle resolved simulations. They use two different markers, queen and worker markers, to handle different data. The queen marker handles the information on the translational and rotational motion of a particle and integrates the force and torque computed at all the worker markers, while the worker marker handles the fluid–particle interaction. However, these works did not pay attention to the load imbalance of disperse phase in flow configurations where the particulate phase is non-uniformly distributed in the fluid domain such as sediment transport. \cite{uhlmann2004simulation} found that the parallel efficiency dropped by 34\% when refine the grid in the simulation of 48 particles settling in an ambient container using 16 processors, which is caused by the uneven distribution of the particles among processors. The load balance of the disperse phase can strongly affect the parallel efficiency of the particle resolved simulations and deserves great attention.}


{In the present study, sediment transport over an erodible sediment bed is studied by particle resolved simulations with a hybrid parallel approach. To overcome the challenges of load imbalance in the traditional domain decomposition method when encountering highly uneven distributions of particles in sediment transport, the parallel approach of \cite{darmana2006parallelization} originally developed for point particle simulations is modified and implemented into particle resolved simulations. The hybrid parallel approach improves the load balance of the particles where they are extremely non-uniform distributed in the fluid domain (i.e., concentrated at the bottom of the fluid domain), which limits the computation efficiency of the IB method.} The memory requirement of the hybrid approach is reduced by a novel memory optimization technique for spherical particles with equal size. The accuracy of our parallel PRDNS code has been rigorously verified by several benchmark cases. The parallel performance is tested by a challenging sediment transport case with a million spherical particles. Furthermore, a turbulent flow over an erodible sediment bed is simulated. An extraction method is proposed to distinguish the saltating and rolling particles and extract impact and rebound information of the particle-mobile bed interaction. The probability distribution functions (PDF) of several important saltation parameters such as velocity, angle, and spanwise angular velocity of the impact and rebound particles are presented. The splash functions are established for the particle-mobile bed interaction in the turbulent flow, which was rarely investigated in the experiments.


\textcolor{black}{The paper is arranged as follows. The numerical schemes employed in this work and the validation cases are introduced in \S 2. The proposed hybrid parallel approach, memory optimization, and the test of the parallel performance are introduced in detail in \S 3. The turbulent flow over an erodible sediment bed is investigated in \S 4. The final conclusion of the paper is drawn in \S 5.}

\section{Numerical schemes}

In this section, we introduce the governing equations of the particle-laden turbulent flow, the fluid-particle coupling model, the hydrodynamic force and torque model and the collision model. These models has been well documented in the literature, so we will not go into detail here. In addition, several benchmark cases are tested to demonstrate the accuracy of the code.

\subsection{Governing equations}

The particle-laden flow considered here is governed by the Navier-Stokes equations for the carrier phase and Newton-Euler equations for the disperse particulate phase. The motion of an incompressible, Newtonian fluid flow is governed by the following continuity and momentum equations:
\begin{equation}
\nabla\cdot \bm u=0, 
\label{eqn:eq1}
\end{equation}
\begin{equation}
\frac{\partial \bm u}{\partial t}= -\nabla\cdot\left(\bm {uu}\right)-\frac{1}{\rho_f}\nabla p+\nu_f\nabla^2 \bm u+\bm f,
\label{eqn:eq2}
\end{equation}
where $\bm u$ is the fluid velocity, $p$ is the dynamic pressure, $\bm f$ is the volume force, $\rho_f$ is the fluid density, and $\nu_f$ is the fluid kinematic viscosity.

For the numerical discretization of equations (\ref{eqn:eq1}-\ref{eqn:eq2}), a second-order central difference scheme is used for spatial discretization, and a second-order Runge-Kutta (RK2) method is used for fluid time advancement \cite[]{yang2017,yang2018direct,cui2018sharp}. At each substep of the RK2 method, the fractional-step method of \cite{kim1985application} is applied to ensure that the flow velocity is divergence free. The discretized equations for each Runge-Kutta substep are written as follows:
\begin{align}
&do\ k=1,2 \nonumber \\
& \ \ \hat{\bm u}^{k}=\bm u^{k-1}+\Delta t_f\left [ \alpha _{k} \bm H^{k-1}-\beta _{k}\left ( \bm H^{k-2}-\frac{1}{\rho_f }\nabla p^{k-2}\right )\right ]\\
& \ \ \tilde{\bm u}^{k}=\hat{\bm u}^{k}+\alpha_k\Delta t_f\bm f^{k}\\
& \ \ \nabla^2 p^{k-1}=\frac{\rho_f}{\alpha_k \Delta t_f} \nabla \cdot \tilde{\bm u}^{k} \label{eqn:eq5}\\
& \ \ \bm u^{k}=\tilde{\bm u}^{k}-\frac{\alpha_k \Delta t_f}{\rho_f}\nabla p^{k-1}\\
&enddo \nonumber
\end{align}
In the above equations, the superscript $k$ is the index of the Runge-Kutta substep. The coefficients in the RK2 scheme are $\alpha_1=1$, $\beta_1=0$, and $\alpha_2=\beta_2=0.5$. $\Delta t_f$ is the time step for the fluid solver, $\hat{\bm u}^k$ and $\tilde{\bm u}^k$ are the intermediate fluid velocities, $\bm H^k=-\nabla\cdot\left(\bm u^k \bm u^k\right)+\nu_f\nabla^2 \bm u^k$ is the summation of the convection and viscous terms, and $f^k$ is the volume force from particles computed by the direct-forcing immersed boundary method \cite[]{uhlmann2005immersed,luo2007full,kempe2012improved,breugem2012second}.

The translational and angular velocities of a particle are solved by the Newton-Euler equations. For a spherical particle, the equations reduce to
\begin{equation}
\rho_p V_p\frac{d \bm u_p}{d t}=\rho_f\oint_{\partial V}{\bm \tau\cdot \bm n_p d A}+\left(\rho_p-\rho_f\right)V_p\bm g+\bm F_{c,p},
\label{eqn:eq7}
\end{equation}
\begin{equation}
I_p\frac{d\bm \omega_p}{d t}=\rho_f\oint_{\partial V}{\bm r\times(\bm \tau\cdot \bm n_p)d A}+\bm T_{c,p},
\label{eqn:eq8}
\end{equation}
where the subscript $p$ indicates the quantities of particle $p$, $\rho_p$ is the particle density, $V_p$ is the volume of the particle and equals $(4/3)\pi R_p^3$ for a spherical particle with radius $R_p$, $I_p$ is the moment of inertia of the particle and equals $(2/5)\rho_p V_p R_p^2$ for a spherical particle, $\bm u_p$ and $\bm \omega_p$ are the translational and angular velocities of the particle, respectively, $\bm \tau=-p \bm I+\mu_f (\nabla \bm u+\nabla \bm u^T )$ is the hydrodynamic stress tensor (the superscript $T$ indicates transposition of a tensor), $\bm n_p$ is the outward-pointing unit normal vector at the particle surface denoted by $\partial V$, $\bm r$ is the position vector of the particle surface relative to the particle center, $\bm g$ is the gravitational acceleration, and $\bm F_{c,p}$ and $\bm T_{c,p}$ are the collision force and torque acting on the particle, respectively. \textcolor{black}{$\bm F_{c,p}$ and $\bm T_{c,p}$ are computed as following,}
\begin{equation}
\bm F_{c,p}= \sum_{p,q \neq p}^{N_{p} } \left ( \bm F_{n,pq}^{lub}+\bm F_{n,pq}^{col}+\bm F_{t,pq}^{col}   \right )+\bm F_{n,pw}^{lub}+\bm F_{n,pw}^{col}+\bm F_{t,pw}^{col}, 
\label{eqn:force}
\end{equation}
\begin{equation}
\bm T_{c,p}= \sum_{p,q \neq p}^{N_{p} }  R_{p} \bm n_{pq} \times \bm F_{t,pq}^{col}+R_{p} \bm n_{pw} \times \bm F_{t,pw}^{col}, 
\label{eqn:torque}
\end{equation}
\textcolor{black}{where $N_{p}$ is the particle number, $\bm F^{lub}$ is the lubrication force, $\bm F^{col}$ is the collision force, $\bm n$ is the normal unit vector of contact, the subscript $n$ and $t$ indicates the normal and tangential direction, respectively, the subscript $pq$ and $pw$ indicates the contact between particle $p$ with particle $q$ and wall, respectively. The two-parameter lubrication model proposed by \cite{costa2015collision} is used to account for the lubrication force. The adaptive collision time model (ACTM) \cite[]{kempe2012collision,biegert2017collision} is employed to account for the collision force in the normal and tangential directions. The stiffness and damping coefficient in this model is adaptively calibrated based on the collision time $T_c=N\Delta t$, where $N=10$ is employed following the suggestion of \cite{kempe2012collision}. To resolve the drastic changes in particle velocity during collisions, the particle time substep $\Delta t_p=\Delta t_f/50$ is used in the following simulations \cite[]{darmana2006parallelization,deen2007review,capecelatro2013euler,finn2016particle}.}

\subsection{Validation}

In this part, we validate the accuracy of the immersed boundary method and collision model step by step. Three cases are simulated and compared with numerical or experimental results in the literature.

\subsubsection{A fixed spherical particle in uniform cross flows}

\begin{figure}[H]
\centering
\includegraphics[width=10cm]{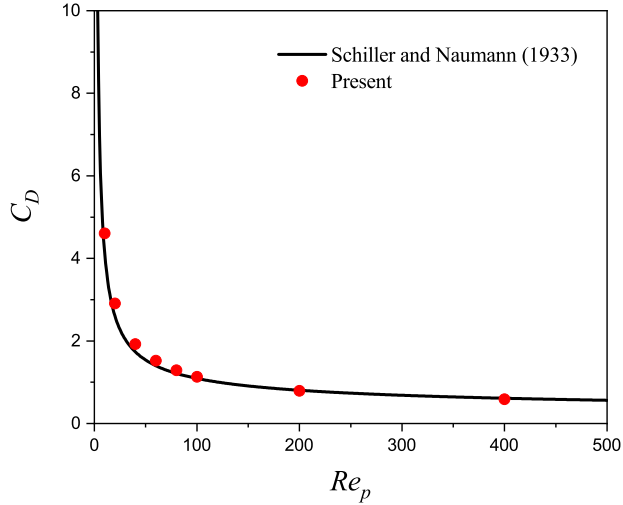}
\caption{Variation of drag coefficient with the particle Reynolds number for a spherical particle in uniform cross flows.}
\label{fig:drag}
\end{figure}

To validate the accuracy of the IB method, a fixed spherical particle in uniform cross flows is simulated. A spherical particle is placed at the center of the domain. The dimensions of the domain are $15D_p\times 6D_p\times 6D_p$ resolved by a $300\times 120\times 120$ grid (20 grid points in $D_p$) along the streamwise, vertical, and spanwise directions, where $D_p$ is the diameter of the particle. Inflow and outflow boundary conditions are implemented at the inlet and outlet, respectively, and periodic boundary conditions are enforced at the side boundaries. The computed drag coefficients at different particle Reynolds numbers (defined as $Re_p=U_\infty D_p/\nu_f$, $U_\infty$ is the inflow velocity) are shown in Fig.~\ref{fig:drag}. The solid line is the empirical drag law (the S-N law), $C_D=({24}/{Re_p})(1+0.15Re_p^{0.687})$, which was proposed by \cite{schiller1933grundlegenden}. It is seen from Fig.~\ref{fig:drag} that the computed results are in good agreement with the S-N law, which validates the present implementation of the multi-direct-forcing immersed boundary method.

\subsubsection{Sedimentation of a spherical particle in a quiescent fluid}

\begin{figure}[H]
\centering
\hspace{-16mm}
\includegraphics[width=8cm]{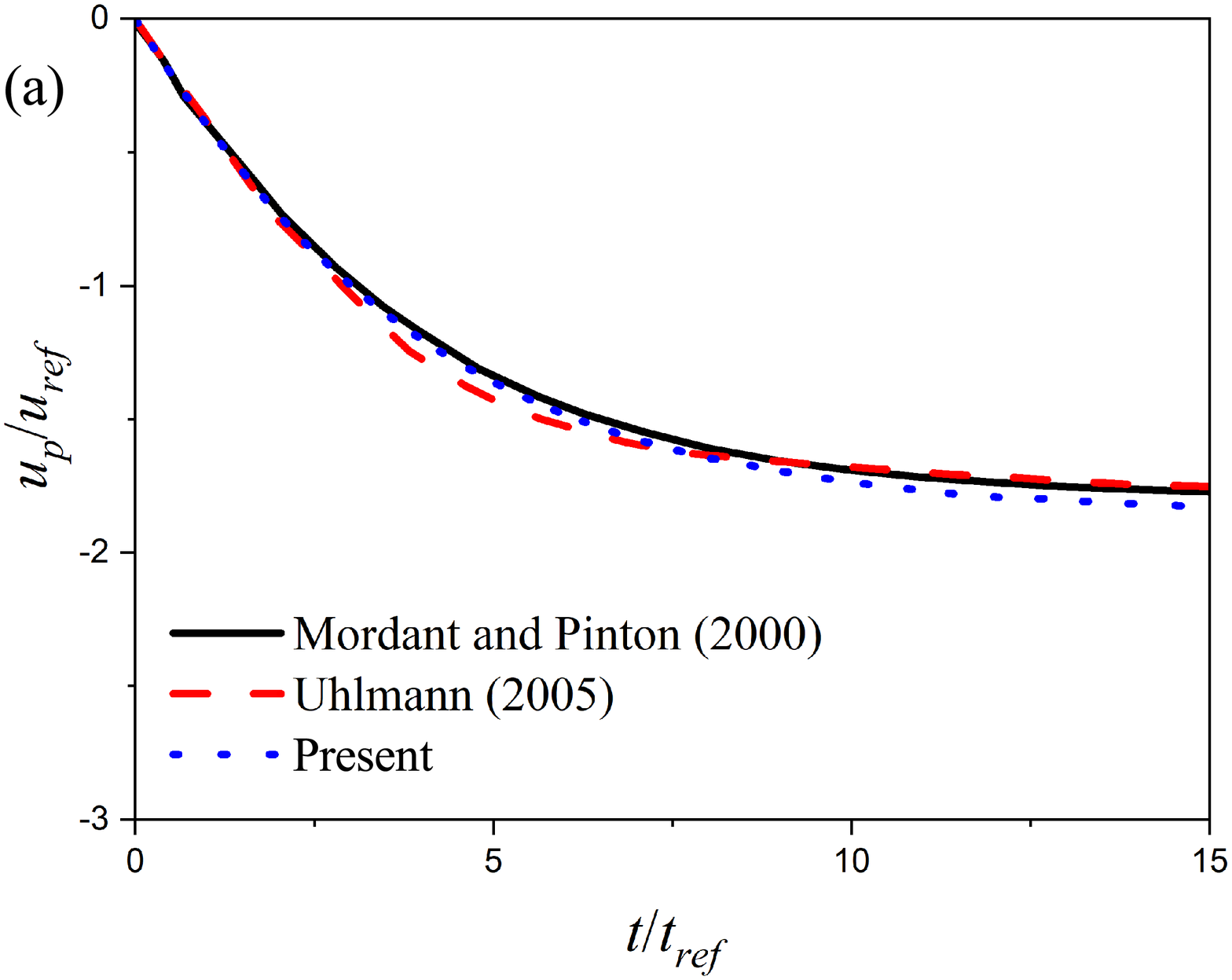}
\hspace{-10mm}
\includegraphics[width=8cm]{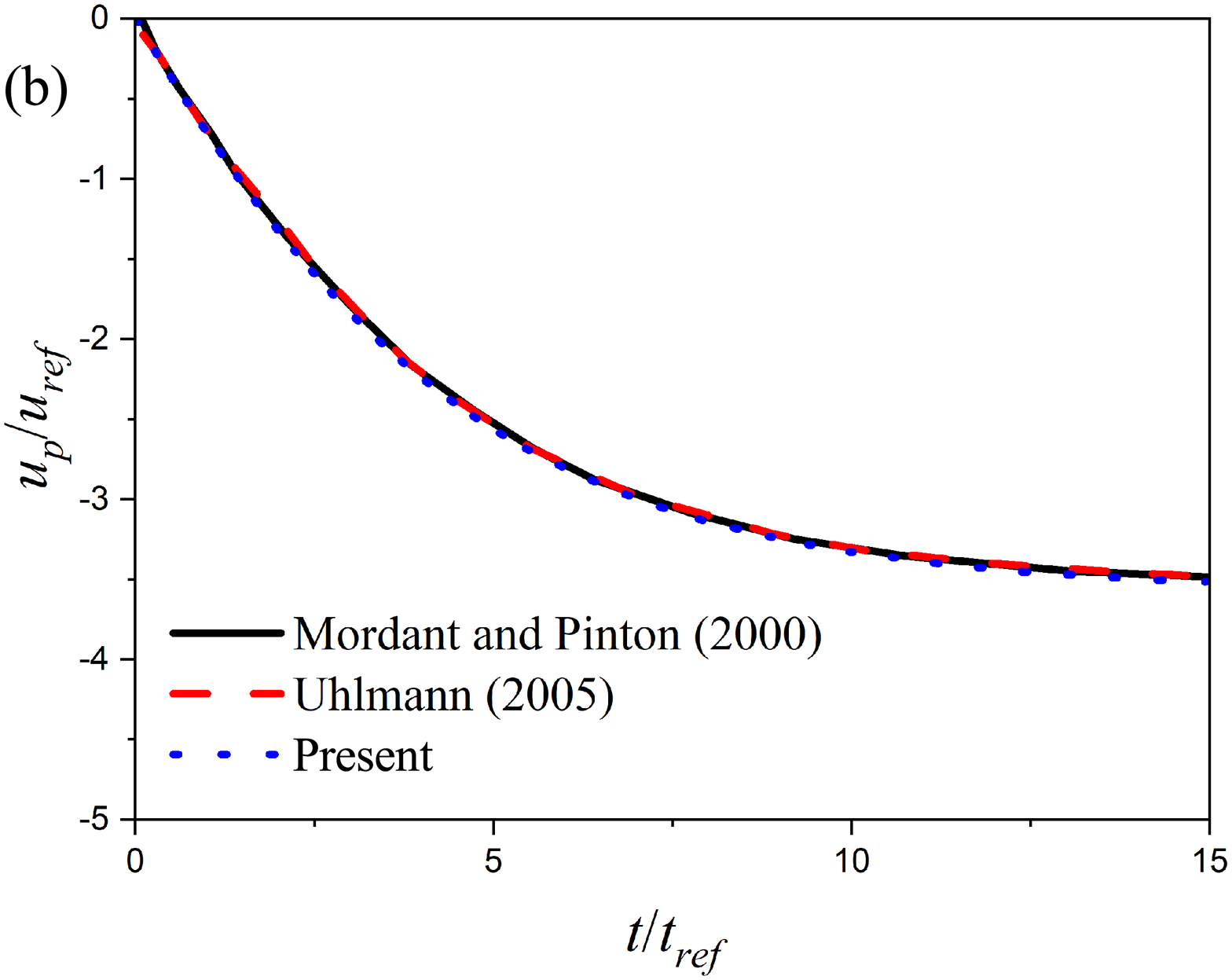}
\hspace{-16mm}
\caption{Comparison of the settling velocity for different density ratios: (a) $\rho_p /  \rho_f=2.56$ and (b) $\rho_p / \rho_f=7.71$. The settling velocity and the time are normalized by $ u_{ref}=\sqrt{D_p\left |g\right |}$ and $t_{ref}=\sqrt{D_p/\left |g\right |}$, respectively.}
\label{fig:settle}
\end{figure}

Next, we validate the accuracy of the IB method for a moving particle using the case of the sedimentation of a spherical particle in a quiescent fluid with different density ratios. The computation configuration is the same as that in \cite{uhlmann2005immersed}, and the parameters used here are listed in Table~\ref{table:4}. Sufficiently large domain size and periodic boundary conditions are applied in all three directions to match the experiment of \cite{mordant2000velocity}. The computational domain size is $L_x\times L_y\times L_z=7.68D_p\times 54D_p\times 7.68D_p$ with a grid of $N_x\times N_y\times N_z=128\times 1024\times 128$ (16.7 grid points in $D_p$). The particle is initially placed at $x=L_x/2$, $y=0.9L_y$, $z=L_z/2$ and released under gravity from rest at $t=0$. The computed settling velocity is displayed in Fig.~\ref{fig:settle}. Compared with the measurement data of \cite{mordant2000velocity} and the simulation results of \cite{uhlmann2005immersed}, it is seen that the present results under different density ratios are all in good agreement with them, validating the current code for moving particles.

\begin{table}[htb]
   \centering
   \caption{Parameters used in the simulation of spherical particle sedimentation.}
   \setlength{\tabcolsep}{3.0mm}{
   \begin{tabular}{cccccccc}
   \toprule
   $Re_p$ & $\rho_p/\rho_f$ & $g$ & $D_p$ & $ \nu_f$  & $D_p/\Delta x$ & $\Delta t$ \\
   377 & 2.56 & 9.81 & 0.167 & $1.04\times 10^{-3}$ & 16.7 & $10^{-3}$ \\
   283 & 7.71 & 9.81 & 0.167 & $2.68\times 10^{-3}$ & 16.7 & $10^{-3}$ \\ \bottomrule
   \end{tabular}}
   \label{table:4}
\end{table}

\subsubsection{Normal particle-wall collision in a viscous fluid}

To validate the accuracy of the collision model, the bouncing motion of a single particle in a viscous fluid with different Stokes numbers is simulated. The computational configuration is the same as that in \cite{biegert2017collision}, and the parameters used in the simulation are listed in Table~\ref{table:5}.

\begin{table}[htb]
   \centering
   \caption{Parameters used in the simulation of the particle-wall collision case.}
   \setlength{\tabcolsep}{3.0mm}{
   \begin{tabular}{ccc}
   \toprule
   $St$ & 27 & 152 \\
   $Re_p$ & 30 & 164 \\
   $D_{p}\ (\text{m})$ & 0.006 & 0.003 \\
   $u_{in}\ (\text{m/s})$ & 0.519 & 0.585 \\
   $\rho_p/ \rho_f$ & 8.083 & 8.342 \\
   $\nu_f\ (\text{m}^2/\text{s})$ & $1.036\times 10^{-4}$ & $1.070\times 10^{-5}$ \\
   $e_n$ & 0.97 & 0.97 \\
   $g\ (\text{m}/\text{s}^2)$ & 9.81 & 9.81 \\
   $L_{x}\times L_{y}\times L_{z}$\ $(\text{m})$& $12.8D_p\times 25.6D_p \times 12.8D_p$ & $6D_p\times 70D_p \times 6D_p$ \\
   Grid number & $256\times 512 \times 256$ & $120\times 1400 \times 120$ \\
   $D_p/\Delta x$ & 20 & 20 \\ \bottomrule
   \end{tabular}}
   \label{table:5}
\end{table}

\begin{figure}[htb]
\centering
\hspace{-16mm}
\includegraphics[width=8cm]{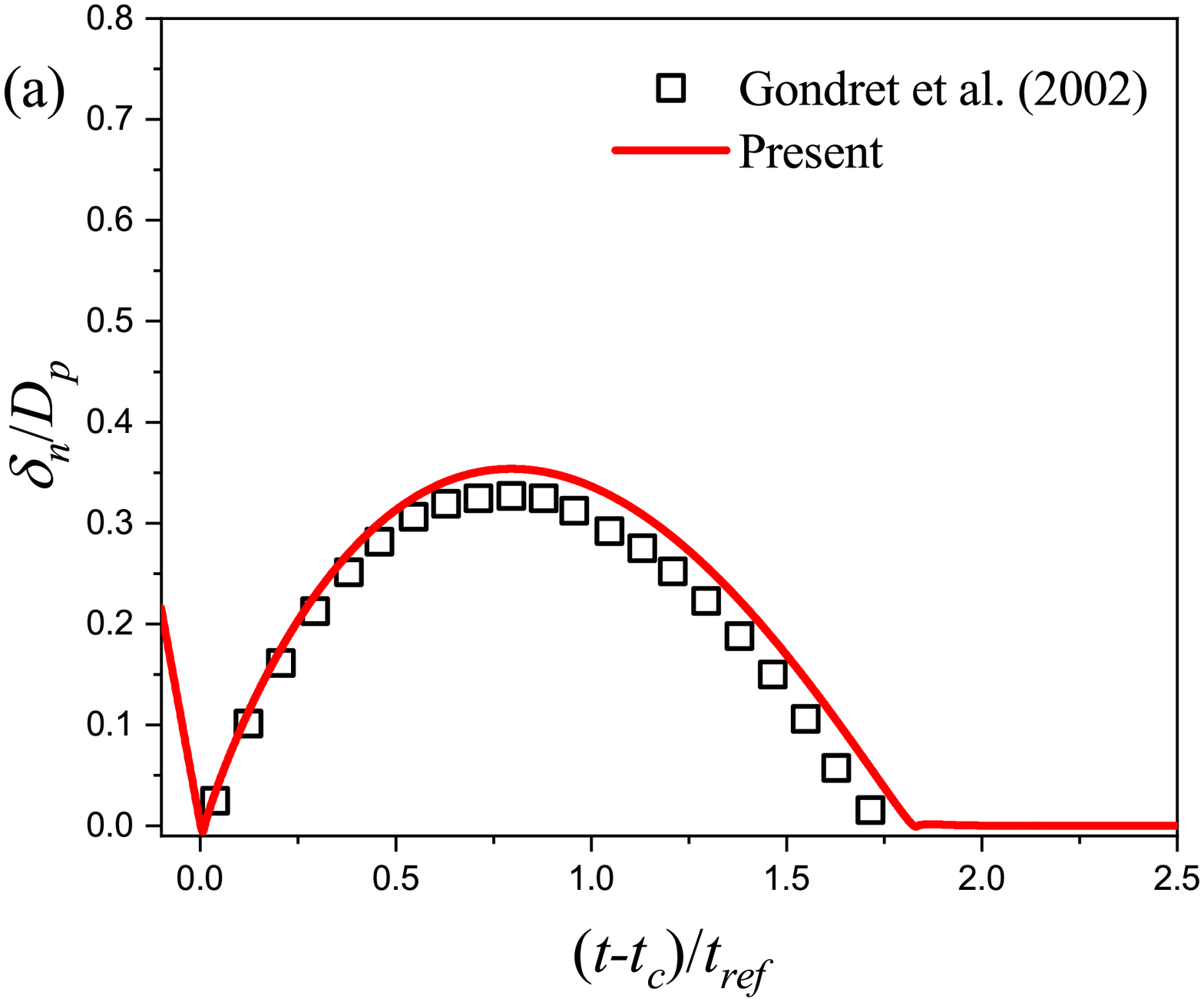}
\hspace{-10mm}
\includegraphics[width=8cm]{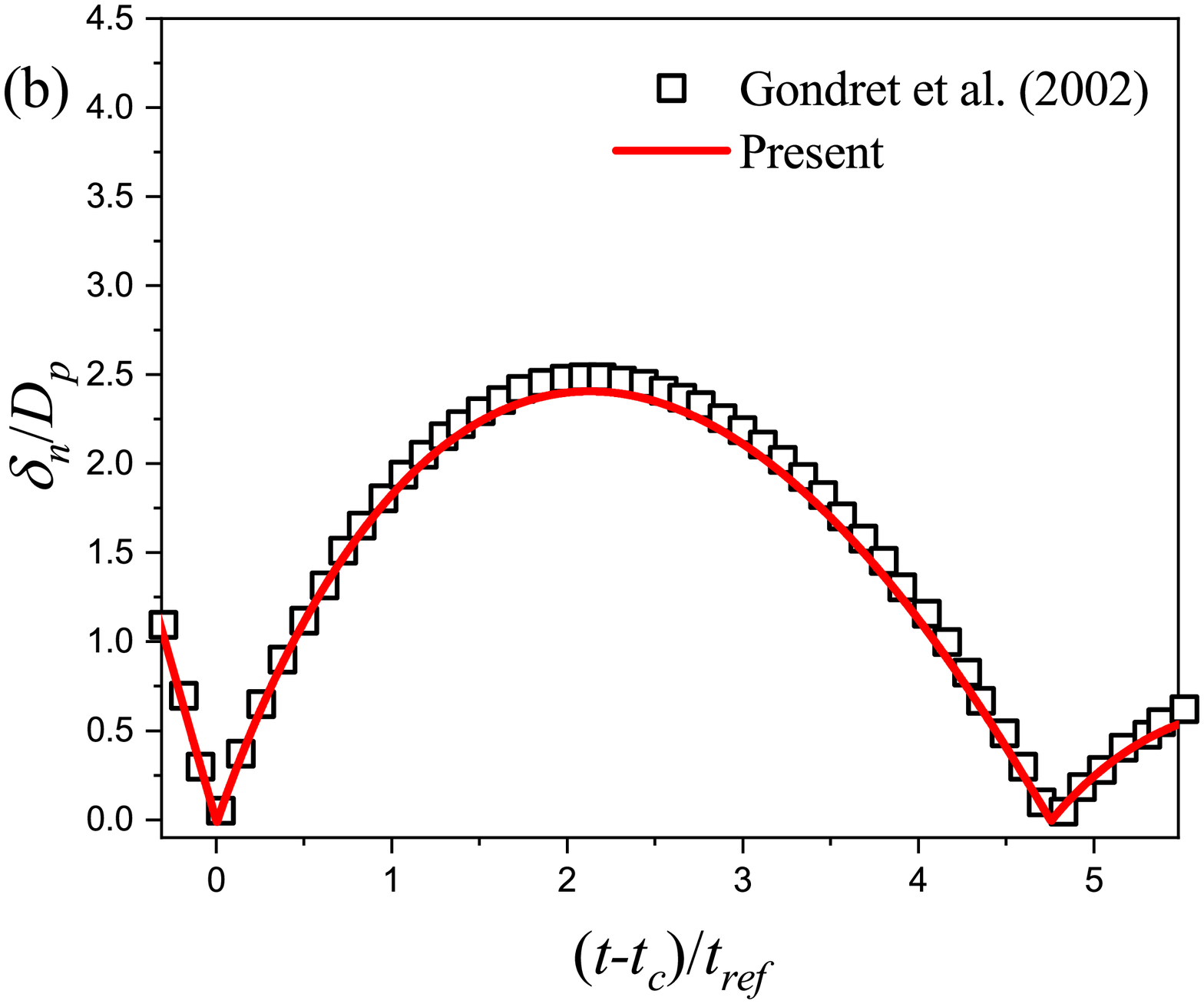}
\hspace{-16mm}
\caption{Comparison of the particle trajectories between the experiment and the present simulation for different Stokes numbers: (a) $St=27$ and (b) $St=152$.}
\label{fig:collision}
\end{figure}

A periodic boundary condition is imposed in the streamwise and spanwise directions, and a no-slip boundary condition is imposed on both the top and bottom surfaces. The particle is initially placed at $x=L_x/2$, $y=L_y-0.75D_p$, $z=L_z/2$. We prescribe the falling velocity of the particle following \cite{biegert2017collision}, that is, accelerate it smoothly and let $u_{in}$ match the Stokes number in the experiment \cite[]{gondret2002bouncing} before the collision, as
\begin{equation}
   u_p(t)=u_{in}\left ( e^{-40t}-1  \right ),\quad \text{if} \quad \delta _{n}>R_{p}.
\end{equation}
Once the particle reaches a distance of $\delta _{n}=R_{p}$, we turn off the prescribed velocity. Then, the particle moves under hydrodynamic, gravitational, buoyant, and collision forces. Figure~\ref{fig:collision} shows the computed particle trajectories at different Stokes numbers. The results are in good agreement with the experiment of \cite{gondret2002bouncing}.

To summarize section 2.2, we conducted simulations of three test cases to validate the IB method for stationary and moving particle problems and the particle collision model. The results are in good agreement with the experimental and simulation data reported in the literature, confirming the accuracy of the present particle-resolved direct numerical simulation code. \textcolor{black}{In the next section, we will introduce the hybrid parallel approach for improving the load balance in the PRDNS of sediment transport.}

\section{Hybrid parallel approach and memory optimization}
\subsection{Parallel approach for the carrier phase}
\begin{figure}[H]
\centering
\includegraphics[width=10cm]{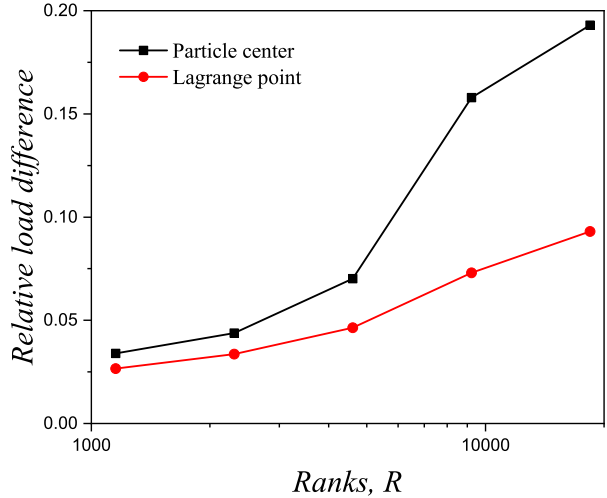}
\caption{The relative load difference for the particle centers and the Lagrangian points on the particle surface of sediment transport case in section 3.5.}
\label{fig:load_diff}
\end{figure}

\textcolor{black}{The carrier phase is parallelized by the domain decomposition method \cite[]{uhlmann2004simulation,tsuji2008spontaneous,gopalakrishnan2013development,wang2017parallel,pozzetti2019parallel,dufresne2020massively}. For sediment transport as shown in Fig.~\ref{fig:sketch_efficiency}, particles are concentrated at the bottom of the channel. Hence, the domain is only decomposed in the horizontal plane to avoid the significant load difference in the vertical direction. Although it can improve the load balance of particles, the load difference still exists because the distribution of particles in the horizontal plane is also uneven in sediment transport. If the disperse phase of sediment transport case in section 3.5 is also parallelized by the domain decomposition method, the relative load difference for the particles is shown in Fig.~\ref{fig:load_diff}. The relative load difference for the particle center (or the Lagrangian point) is computed by the maximum value of $(N_{p,i}-N_{p,ave})/N_{p,ave}$ (or $(N_{l,i}-N_{l,ave})/N_{l,ave}$) in all ranks, where $N_{p,i}$ and $N_{l,i}$ are the particle centers and the Lagrangian points in the subdomain of rank $i$, and $N_{p,ave}$ and $N_{l,ave}$ are the average number of the particle centers and the Lagrangian points in all ranks. It represents the increased computational time due to the load difference compared with the ideal load. The relative load difference increases as the number of ranks increases. It can reach 16\% for the particle center and 7\% for the Lagrangian point when the decomposed subdomain size is $2D_p\times4D_p$ with 9216 ranks, which may impact the parallel efficiency. Therefore, it may be helpful to improve the load balance beyond the domain decomposition method for the disperse phase and interphase coupling in the PRDNS of sediment transport, which will be introduced below.}

\subsection{Parallel approach for the disperse phase}
\begin{figure}[htb]
\centering
\includegraphics[width=0.8\textwidth]{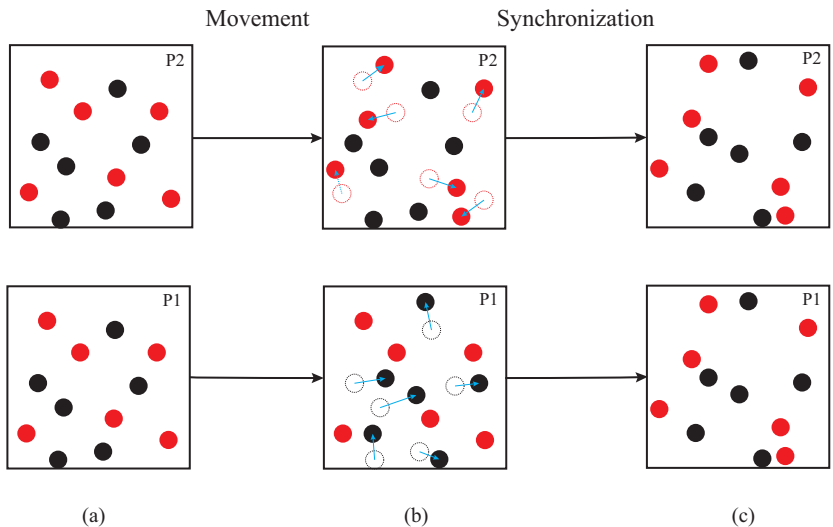}
\caption{Illustration of the mirror domain technique for the disperse phase. (a) Initial state of the particles in each processor; (b) particle movement in each processor; and (c) final state of the particles after synchronization in each processor. The black and red particles belong to different particle subsets in P1 and P2, respectively. The blue arrow is the particle movement path.}
\label{fig:mirror}
\end{figure}

To balance the workload among different ranks, the disperse phase is parallelized by the mirror domain technique \cite[]{darmana2006parallelization}. Different from the domain decomposition method where each processor corresponds to a specific subdomain and transmits data at the subdomain boundaries, each processor in the mirror domain method stores the same total particle data but only deals with a subset of them.
Figure \ref{fig:mirror} illustrates the mirror domain technique, where P1 and P2 denote two processors. Each processor stores the data of all particles. The initial states of all particles in the two processors are the same, as shown in Fig.~\ref{fig:mirror}a. At the next time step, when the particles move along the arrows to the new positions, as shown in Fig.~\ref{fig:mirror}b, processor P1 only deals with the particles in black, while processor P2 deals with the particles in red.
After the computation of particle movements, synchronization is performed to update the data of all particles in each processor. Then, the same total particle data are obtained and prepared for the next step of computation, as shown in Fig.~\ref{fig:mirror}c. The main advantage of the mirror domain technique compared with the domain decomposition technique is that the number of particles stored and computed in each processor is the same, regardless of the particle distribution, which can help achieve excellent load balance for the disperse phase. However, it should be noted that the number of particles is limited by the storage size of each processor.

\subsection{Parallel approach for interphase coupling}

\textcolor{black}{The interphase coupling between carrier phase and disperse phase is realized by the direct-forcing IB method \cite[]{uhlmann2005immersed,luo2007full,kempe2012improved,breugem2012second}. The information of the fluid Eulerian grids and the particle Lagrangian points is coupled by interpolation and diffusion operations through a regularized Dirac delta function $\delta_d$. In the mirror domain technique, the particle information is needed in the IB method, such as the particle position $\bm x_p$, translational velocity $\bm u_p$, and angular velocity $\bm \omega_p$ of all the particles, which should be stored and synchronized in each rank. Therefore, we only need to map particles (i.e., find all particles that affect the rank's subdomain, and this rank handles all Lagrangian points of these particles) to each ranks when calculating the IB force. As shown in Fig.~\ref{fig:load_diff}, the load balance of the Lagrangian point is better than the particle center, especially when a large number of ranks are invoked. Therefore, when calculating the IB force, the Lagrangian points of all particles are mapped to each rank rather than particle center to improve the load balance in the calculation of interphase coupling.}

\textcolor{black}{The original mirror domain technique of \cite{darmana2006parallelization} is developed for point-particle simulations. The present work modifies it so that it can be applied in the particle resolved simulations. The modifications are in the calculation of interphase coupling:}

\begin{figure}[H]
\centering
\includegraphics[width=0.8\textwidth]{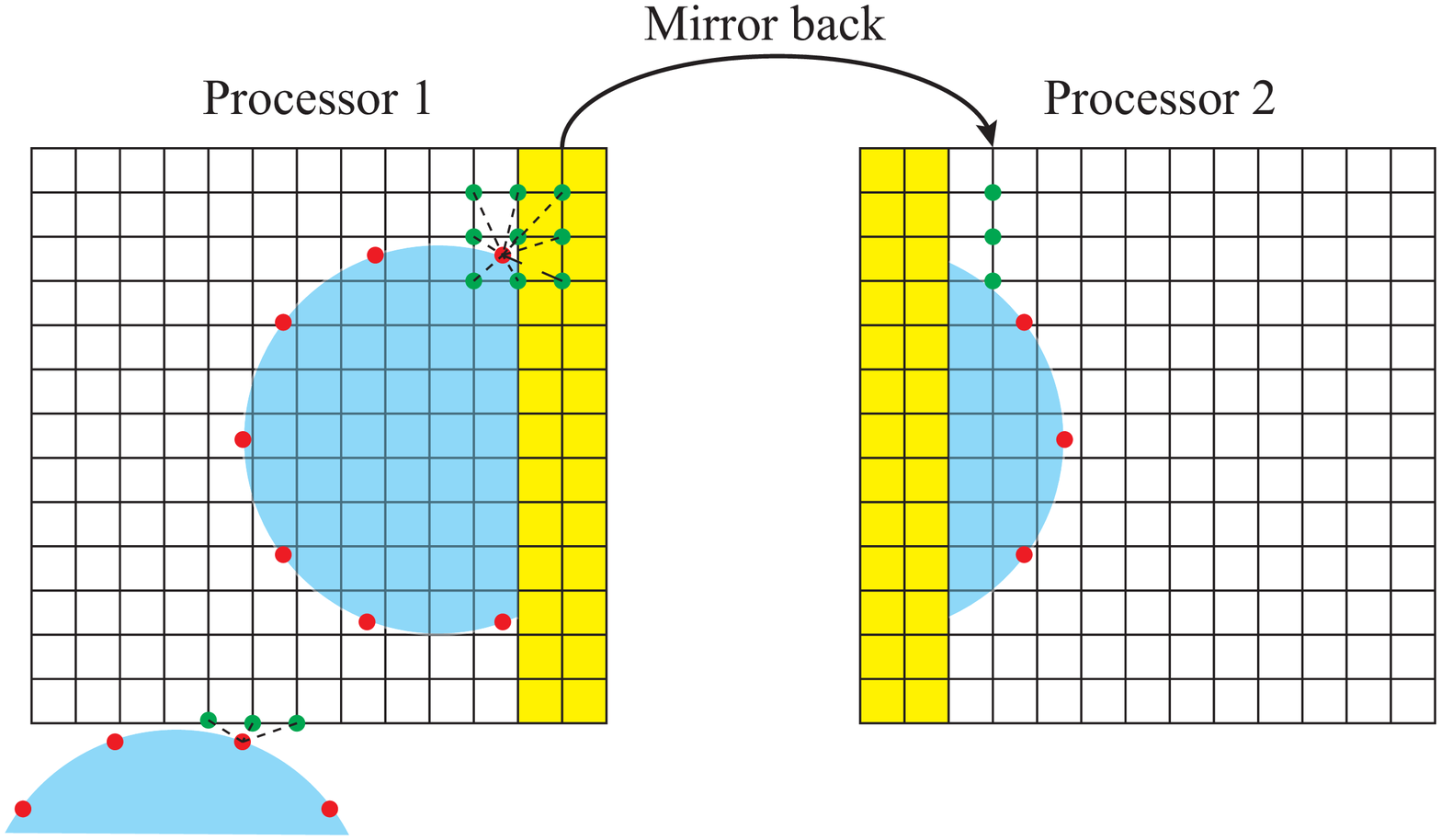}
\caption{Two-dimensional illustration of data transmission in particle-fluid interaction calculations. The blue circle is the finite-sized particle. The yellow region is the ghost-cell region. The red points are the Lagrangian points on the particle surface. The green points are the Eulerian grid points affected by a Lagrangian point.}
\label{fig:interaction}
\end{figure}

(1) In a particle resolved simulation, a particle is treated as having a finite size by using Lagrangian points on the particle surface (the red points in Figure~\ref{fig:interaction}) rather than as a sizeless point in a point-particle simulation. Because the carrier phase and disperse phase are parallelized with different methods, the particle position data need to be mirrored to each fluid subdomain. For a point-particle simulation, the subdomain in which a particle is mirrored is unique, while this may not be the case for a particle resolved simulation in which the position of the Lagrangian points may be mirrored to multiple subdomains. For example, for the particle on the bottom left of Fig.~\ref{fig:interaction}, even though the particle center is not in the subdomain, as shown, the particle position data still need to be mirrored to this subdomain, as there are two Lagrangian points in it.

(2) Since the finite size particles are decomposed to multiple parts and located in different subdomains when the particle is near the subdomain boundary, as shown in Fig.~\ref{fig:interaction}, the fluid force on the particle is calculated by summing the forces in all these parts. Therefore, the Lagrangian force data transmission from different fluid subdomains is needed, which does not need in the point-particle simulation.

(3) The hydrodynamic force of a particle acting on the fluid is calculated by diffusing the Lagrangian forces to the surrounding Eulerian grid cells (the green points in Fig.~\ref{fig:interaction}).
The Lagrangian forces affect not only the Eulerian grid cell where the Lagrangian point is located but also the surrounding Eulerian grid cells due to the diffusion operation. For example, as shown in Fig.~\ref{fig:interaction}, a diffused Lagrangian force may exist in the ghost-cell region (the yellow region in Fig.~\ref{fig:interaction}) if the Lagrangian point is near a subdomain boundary. Thus, it is required that the diffused force in the ghost-cell region be mirrored back to the adjacent subdomain and superimposed with the existing force.

\begin{figure}[H]
\centering
\includegraphics[width=10cm]{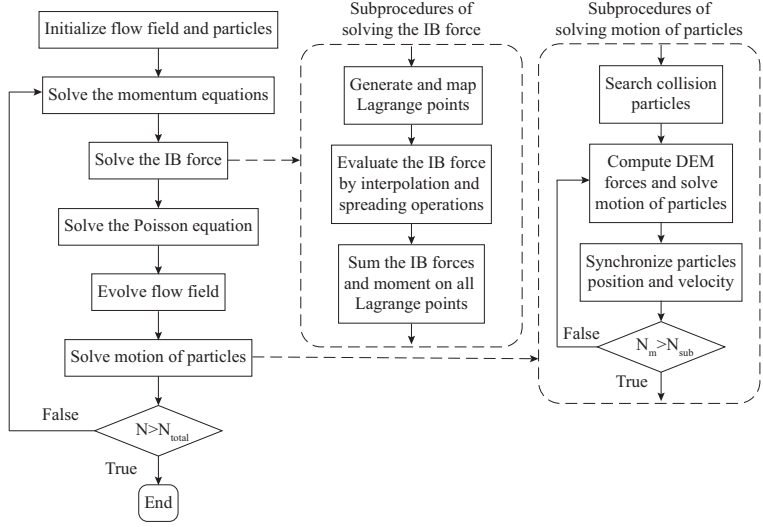}
\caption{The flowchart of the hybrid parallel approach in PRDNS with the direct-forcing immersed boundary method.}
\label{fig:parallel_process}
\end{figure}

\textcolor{black}{By the special consideration in paralleling carrier phase, disperse phase, and interphase coupling, the hybrid parallel approach improves the load balance in the PRDNS of sediment transport. Figure~\ref{fig:parallel_process} shows the flowchart of the hybrid parallel approach. The right dashed frame of Fig.~\ref{fig:parallel_process} elaborates the key subprocedures of solving the IB force and the motion of particles, respectively. In section 3.5, we will study the elapsed clock time of the overall and each subprocedure.}

\subsection{Memory optimization}

The available computer memory on high-performance computing platforms limits the number of particles that can be simulated, especially for the mirror domain technique, in which the data of particle quantities, such as the particle position $\bm x_p$ and translational and angular velocities $\bm u_p$ and $\bm \omega_p$, need to be stored on each computing node. In addition to the above quantities, five relative quantities are also needed in the collision model, including the relative displacements $\bm \delta_n$ and $\bm \delta_t$, the relative velocities $\bm u_{cp,n}$ and $\bm u_{cp,t}$, and the impact velocity $u_{in}$. These quantities are defined in the Appendix A. Although $\bm \delta_n$, $\bm u_{cp,n}$, and $\bm u_{cp,t}$ can be calculated directly from the particle position and velocity without memory consumption, $\bm \delta_t$ and $u_{in}$ require large memories as the particle number increases. The memory consumption for $\delta_t$ and $u_{in}$ is determined by the number of particle pairs, which equals the square of the particle number. Next, we show that $\bm \delta_t$ and $u_{in}$ can be stored with a significantly reduced memory cost.

Here, we propose a novel memory optimization technique to minimize the memory cost of the particle-related variables for spherical particles of the same size. The key idea is to utilize the feature of spherical particles of the same size that each particle can only be in contact with up to twelve surrounding particles in the case of dense packing \cite[]{dai2019modes}. {Generally, the particle information in the 26 surrounding subsets should be stored for relative quantities since one subset could interact at most with the 26 surrounding subsets. But in the memory optimization method, we only store particles that collide, which is at most twelve particles rather than the whole 26 subsets. Without this method, the memory requirement for each relative variable is $N_p^*\times N_p^*\times 26$, where $N_p^*$ is the number of particles residing in each subset. As a result, one can only afford $O(10^4)$ of particles on a common computing platform. By implementing the optimization method developed in this study, the memory cost can be greatly reduced by a factor of $26\times N_p^*/12$ to $12N_p^*$.} As a result, we can handle millions of particles on a common high-performance platform, which is comparable to the highest particle number in the recent particle resolved simulation reported by \cite{kidanemariam2017formation}.

\begin{figure}[H]
\centering
\includegraphics[width=0.8\textwidth]{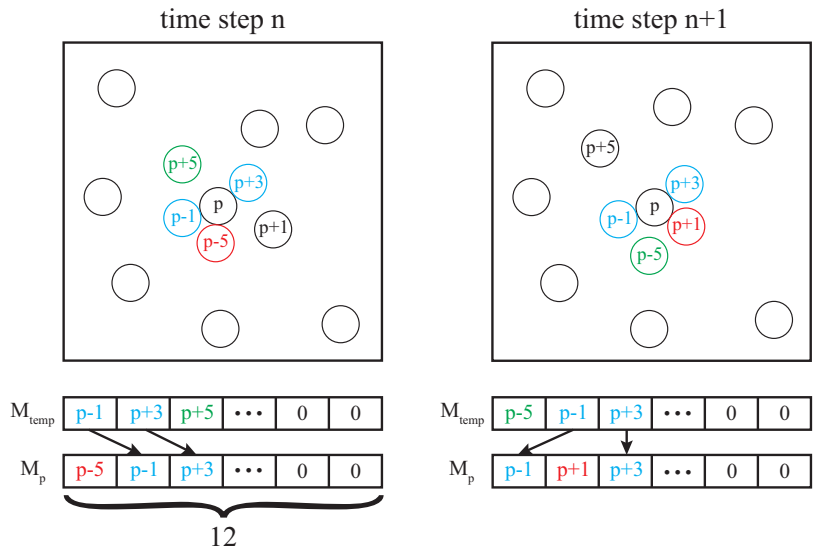}
\caption{Sketch of the memory optimization method with the dense packing concept. $M_p$ and $M_{temp}$ are two arrays to store the collision information at time steps $n$ and $n-1$, respectively. The circles in different colors indicate particles in different collision events.}
\label{fig:memory}
\end{figure}

Furthermore, the particle collision model requires identifying the collision state between two particles, i.e., whether they are undergoing an existing collision event, a new collision event, or a finished collision event, because different collision events have different operations on $u_{in}$ and $\bm \delta_t$. The surrounding particles in contact with a specific particle usually change with time. For example, as shown in Fig.~\ref{fig:memory}, at time step $n$, particle $p$ is in contact with particle $p-5$, particle $p-1$, and particle $p+3$. However, at time step $n+1$, the contact particles of particle $p$ change to particle $p-1$, particle $p+1$, and particle $p+3$. Therefore, additional quantities and treatments are needed to identify and advance the collision state. Here, we use an array $M_p$ to store the contact particle indexes, as well as the $u_{in}$ and $\bm \delta_t$ of particle $p$ at time step $n$, and a temporary array $M_{temp}$ to store the same kind of data at the last time step $n-1$.
Given the above new quantities, three collision events can be identified as follows. (a) If it is a new collision event, that is, a particle collides with particle $p$ at this time step but not at the last time step, then its index is not in $M_{temp}$ (e.g., the red particle indexes $p-5$ and $p+1$ at time steps $n$ and $n+1$, respectively), $u_{in}$ needs to be recorded, and $\bm \delta_t$ needs to be initialized. (b) If it is an existing collision event, i.e., a particle collides with particle $p$ and its index is already in $M_{temp}$ (e.g., the blue particle indexes $p-1$ and $p+3$ at time steps $n$ and $n+1$, respectively), then $u_{in}$ is inherited from $M_{temp}$, and $\bm \delta_t$ is advanced from $M_{temp}$. (c) If a collision event is finished at this time step, i.e., the particle index is in $M_{temp}$ but not $M_p$, then the particle completes the collision (e.g., the green particle indexes $p+5$ and $p-5$ at time steps $n$ and $n+1$, respectively). In this case, $u_{in}$ and $\bm \delta_t$ are reset to zero.

Finally, it is noted that the present memory optimization method can be applied not only in the mirror domain technique but also in other parallel methods or point-particle simulations.

\subsection{Parallel scaling of the hybrid parallel approach}

\begin{figure}[H]
\centering
\includegraphics[width=\textwidth]{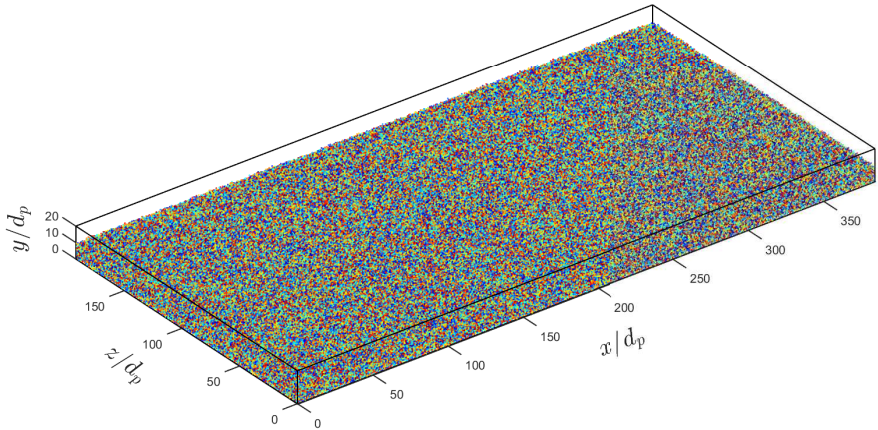}
\caption{The sketch of sediment transport case with over a million sphere particles.}
\label{fig:sketch_efficiency}
\end{figure}

{To investigate the parallel scaling of the hybrid parallel approach,} we conduct a challenging sediment transport simulation for a benchmark case where an erodible sediment bed is comprised by 1050624 spherical particles, as shown in Fig.~\ref{fig:sketch_efficiency}. Particles are deposited at the bottom of the channel, causing an extremely non-uniform distribution of the particles in the vertical direction.

The computational domain is $L_x \times L_y \times L_z=(384\times 20 \times 192)D_p$ on a uniform Cartesian grid of $N_x \times N_y \times N_z=3840\times 200 \times 1920$. The surface of each spherical particle is resolved by 315 Lagrangian points. Other carrier and disperse phase parameters are the same with the case in section 4. {It should be noted that since we generate the Lagrangian points one particle by one particle and compute the IB force, therefore we only dynamically store in total of 315 Lagrangian points in memory.}
{The dimensionless flow time step is $\Delta t_f=10^{-6}$ that makes the CFL number less than 0.5.} The simulations were carried out on the Tianhe-2A supercomputer with 48 to 192 nodes (1152 to 4608 ranks). Each node owns two Intel Xeon E5-2692 cores and 64G memory. The domain is only decomposed in the horizontal plane. The number of ranks in the streamwise direction increases from 24 to 96, while the ranks number remains 48 in the spanwise direction. 

\begin{figure}[H]
\centering
\hspace{-16mm}
\includegraphics[width=8cm]{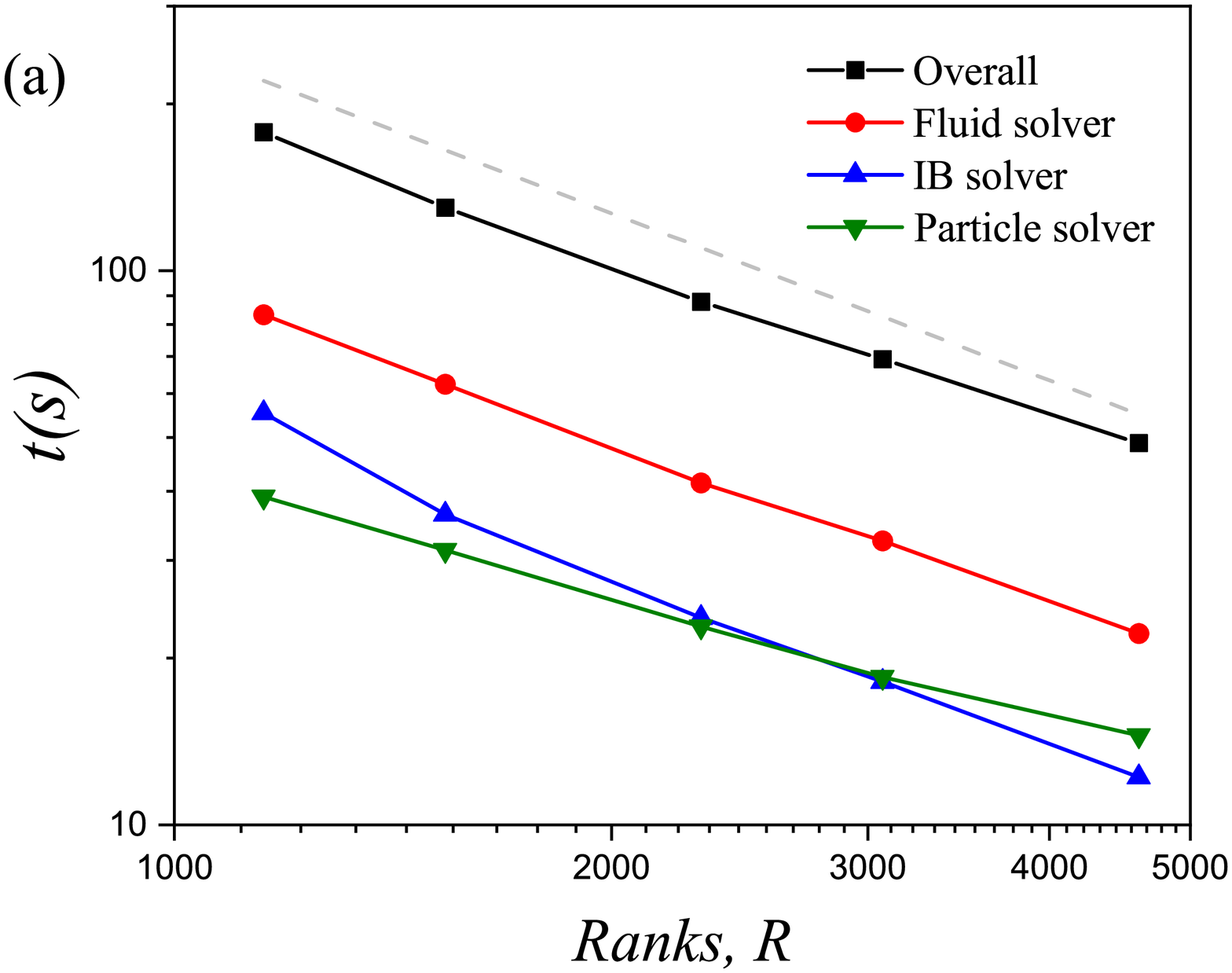}
\hspace{-10mm}
\includegraphics[width=8cm]{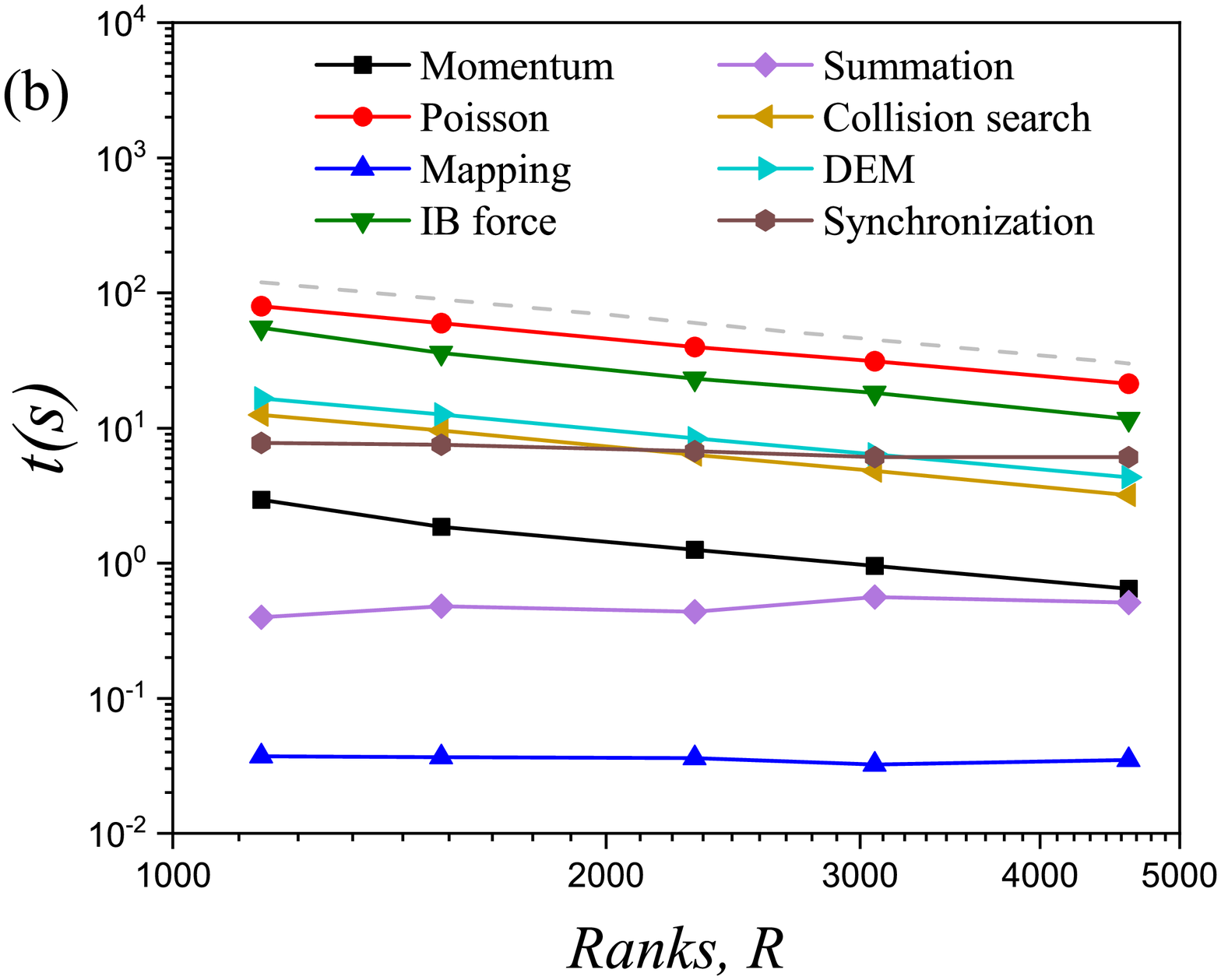}
\hspace{-16mm}
\caption{The elapsed clock time to advance a single fluid time step with MPI rank from 1152 to 4608: (a) timing for overall, fluid, IB, and particle solver and (b) timing for each subprocedure. The gray dash line is the ideal time consumption with the slope $-1.0$.}
\label{fig:rank}
\end{figure}

Figure~\ref{fig:rank} shows the elapsed clock time to advance a single fluid time. The timing is averaged over 100 fluid time steps. As shown in Fig.~\ref{fig:rank}(a), the timing for the fluid, IB, and particle solver achieve good scalability with the slope of $-0.971$, $-1.162$, and $-0.745$, respectively. The slope $s$ is calculated using a linear regression $t=aR^s$. To study the slope difference between different solvers, Fig.~\ref{fig:rank}(b) shows the elapsed clock time for the key subprocedures listed in the flowchart of Fig.~\ref{fig:parallel_process}. And the slopes of them are listed in Table~\ref{table:slope}. The most-consuming subprocedures, including Poisson, IB force, and DEM, have good scalability with a slope close to $-1.0$, which corresponds to the ideal time consumption. However, the time consumption of data communication subprocedures for the disperse phase, including summation, and synchronization remains constant as rank increases. It is because the modified mirror domain method is used to parallel the disperse phase, and the data of disperse phase needs to be transmitted to all the other ranks in these subprocedures. So the amount of communicating data remains constant. The time consumption of the mapping procedure is also constant since all particle needs to determine whether it is located within the rank's subdomain. The time consumption of the mapping and summation subprocedures is negligible in the IB solver. In contrast, the time consumption of the synchronization subprocedure is not negligible in the particle solver. It causes the slope of the particle solver to be less than the IB solver. {The time proportion of the particle synchronization in the total elapsed time is [0.04,0.12] for ranks between 1152 to 4608.} On the whole, the overall timing has good scalability with a slope $-0.966$ which is very close to $-1.0$. And the parallel efficiency $E_n=T_{a}/N_rT_{b}$ is 91\% with $a=1152$ and $b=4608$, where $T_{a}$ and $T_{b}$ are the elapsed time of the parallel code with $a$ and $b$ ranks, respectively, $N_r=b/a$ is the rank ratio. These indicate that the hybrid parallel approach has a good parallel performance for sediment transport simulation with millions of particles.

\begin{table}[htb]
   \centering
   \caption{Slope for the key subprocedures list in the flow chart of Fig.~\ref{fig:parallel_process}.}
   \setlength{\tabcolsep}{20.0mm}{
   \begin{tabular}{cc}
   \toprule
   Subprocedures & Slope \\
   \midrule
   Momentum & -1.1718  \\
   Poisson & -0.9716  \\
   Mapping & -0.0691  \\
   IB force & -1.1765  \\
   Summation & 0.1726  \\
   Collision search & -0.9820  \\
   DEM & -0.9747  \\
   synchronization & -0.2013  \\ \bottomrule
   \end{tabular}}
   \label{table:slope}
\end{table}

\section{Turbulent flow over an erodible sediment bed}

\subsection{Simulation setup}
\begin{figure}[H]
\centering
\includegraphics[width=0.6\textwidth]{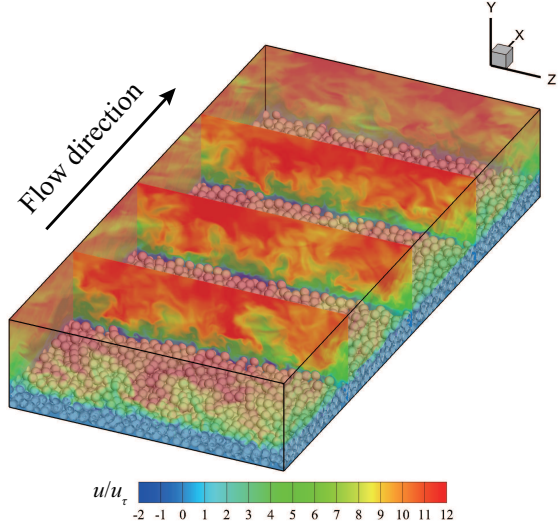}
\caption{Instantaneous snapshot of the turbulent flow over a static sediment bed. The flow field is colored by the nondimensional streamwise velocity $u/u_{\tau}$.}
\label{fig:channel}
\end{figure}

In this section, we simulate the turbulent flow over an erodible sediment bed. The simulation setup is similar to the particle resolved simulation of \cite{ji2014saltation}. The flow is driven by a horizontal body force that is balanced by the shear stress on the sediment bed. The sediment bed consists of $N_p=4608$ particles with two to three layers, as shown in Fig.~\ref{fig:channel}. In the present study, the sediment bed is generated by a sedimentation DEM simulation for particles settling under gravity while turning off the hydrodynamic force, similar to \cite{kidanemariam2014interface}. A periodic boundary condition is imposed in the streamwise and spanwise directions, a no-slip boundary condition is imposed on both the bottom surface and particle surface, and a free-slip boundary condition is imposed on the domain top.

For the carrier phase, the size of the computational domain is $L_x \times L_y \times L_z=(6 \times 1 \times 3)H$ resolved on a uniform Cartesian grid of $N_x \times N_y \times N_z=960 \times 160 \times 480$, where $H$ is the computational domain height in the $y$ direction. Here, $x$, $y$ and $z$ denote the streamwise, wall-normal and spanwise directions, respectively. The Reynolds number is $Re_\tau=u_{\tau}H/\nu=678$, where $u_{\tau}=\sqrt{\tau_b/\rho_f}$ is the friction velocity, $\tau_b$ is the mean shear stress at an effective sediment-bed height $y_b$ (also called the zero-plane displacement), $H_e=H-y_b$ is the effective boundary-layer height, and the superscript $+$ indicates quantities normalized in viscous units (by $u_\tau$ and $\nu_f$). The effective bed height $y_b$ and equivalent bed roughness $k_s$ are determined by fitting the mean velocity profile of the rough bed flow to the classical logarithmic law \cite[]{raupach1991rough,jimenez2004turbulent,singh2007numerical,ji2014saltation,chung2021predicting,kadivar2021review}, which yields $y_b=0.228$ and $k_s=0.224$, which are slightly smaller than the values of $y_b=0.252$ and $k_s=0.242$ in \cite{ji2014saltation}, likely due to the different distribution of particles in the sediment bed. The roughness Reynolds number is $k_s^+=224$, which indicates a fully rough flow regime. The dimensionless time step for the carrier phase is $\Delta t_f=1.5\times 10^{-4}$.

The parameters for the disperse phase are as follows. The density ratio between the particles and fluid is $\rho_p/\rho_f=2.65$. The dimensionless particle diameter is $D_p/H=0.1$. The particle diameter in wall units is $D_p^+= D_p/(\nu_f/u_\tau) = 88$. The mean volume fraction of the disperse phase is 13.4\%. For the parameters in the collision model, the normal restitution coefficient is $e_{n,d}=0.97$, the tangential restitution coefficient is $e_{t,d}=0.39$ and the friction coefficient is $\mu _c=0.15$, which are chosen according to the material properties of the sand particle \cite[]{joseph2004oblique}. The collision time uses $T_c=10\Delta t_f$ following the suggestion by \cite{kempe2012collision}, and the dimensionless time step for the particle is $\Delta t_p=\Delta t_f/50$. The Shields number is ${\it \Theta} =u_{\tau}^2/((\rho_p / \rho_f- 1)gD_p)=0.065$, and the corresponding Galileo number is $Ga=\sqrt{(\rho_p/ \rho_f-1)gD_p^3}/\nu =345$.

It should be noted that while the simulation model and parameters of the present simulation are matched to those of \cite{ji2014saltation}, there is still a difference in the collision model. We adopt the ACTM \cite[]{kempe2012collision,biegert2017collision} while \cite{ji2014saltation} Ji employed the combined finite-discrete element method \cite[]{munjiza1995combined,munjiza2000penalty}. In the ACTM, contact forces are modeled by a spring-dashpot system. In the combined finite-discrete element method, contact forces are calculated by the deformation of particles, which is simulated by the finite-element method. This difference may cause different particle distributions in the sediment bed. {And although the grid resolution $D_p/\Delta x=16$ is enough to resolve the spherical particle, it is insufficient for the small scale turbulent vortices that are smaller than the grid size $\Delta y^+=5.5$. The unresolved small scale turbulent vortices would influence the results. But the degree of the influence on the fluid-sediment interaction is still unclear and requires further research in the future. Hence, the simulation is not focused on providing high-quality quantitative results but on reproducing the main physics underlying the fluid-sediment interaction, which allows establishing the saltation motion of sediment particles.}

\subsection{Statistics verification of the carrier and disperse phases}
Two cases are considered in the present study. The first case is the turbulent flow over a static sediment bed in which all particles are stationary. The second case is the turbulent flow over a mobile sediment bed in which particles are moved under the actions of hydrodynamic, gravitational and collision forces. To make a comparison with \cite{ji2014saltation}, the wall-normal coordinate $Y=y-y_b$ is adopted and normalized by $H_e=H-y_b$ to eliminate the influence of sediment bed height. The effective height of the sediment bed is shown by the gray line in the following figures. The averaging operations for the variables of the carrier and disperse phases are defined in Appendix B.

Figure~\ref{fig:mean velocity} shows the mean streamwise velocity profile. The operator $\left \langle \cdot \right \rangle$ indicates an average over the $x-z$ plane and time. For both the static and mobile cases, our results are generally in good agreement with those of \cite{ji2014saltation}. Above the bed surface, the mean flow velocity in the mobile case is smaller than that in the static case. This is because the entrained particles exert a drag force on the flow and retard the flow accordingly.

\begin{figure}[H]
\centering
\includegraphics[width=10cm]{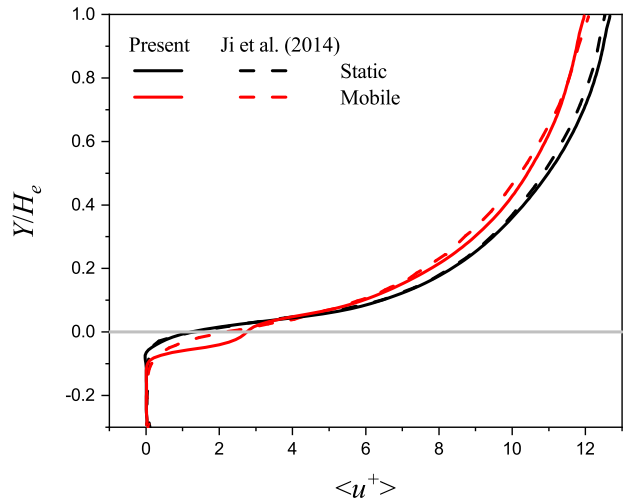}
\caption{Mean streamwise velocity profile of the carrier phase. {The gray line indicates the effective height of the sediment bed. }}
\label{fig:mean velocity}
\end{figure}

\begin{figure}[H]
\centering
\hspace{-16mm}
\includegraphics[width=8cm]{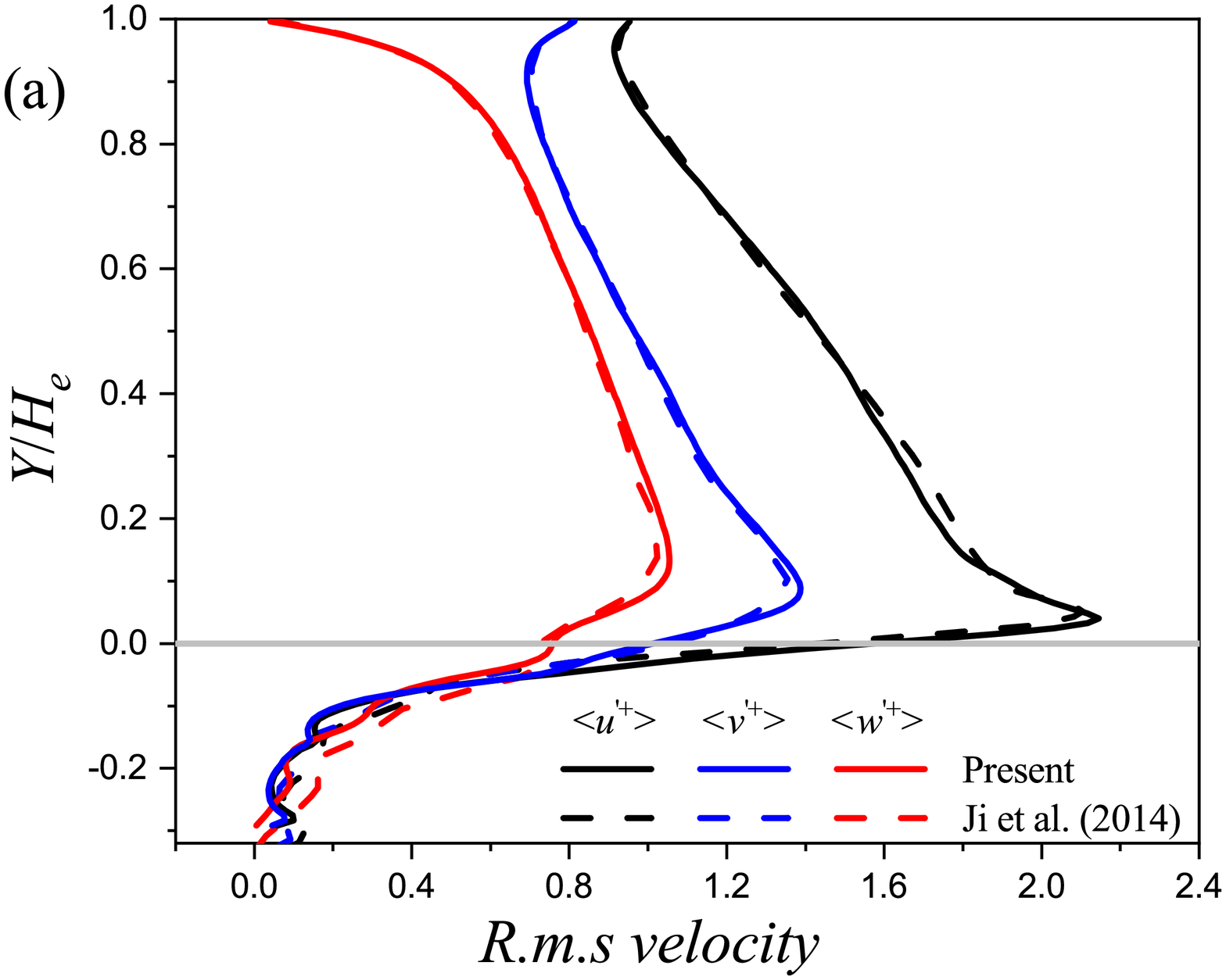}
\hspace{-10mm}
\includegraphics[width=8cm]{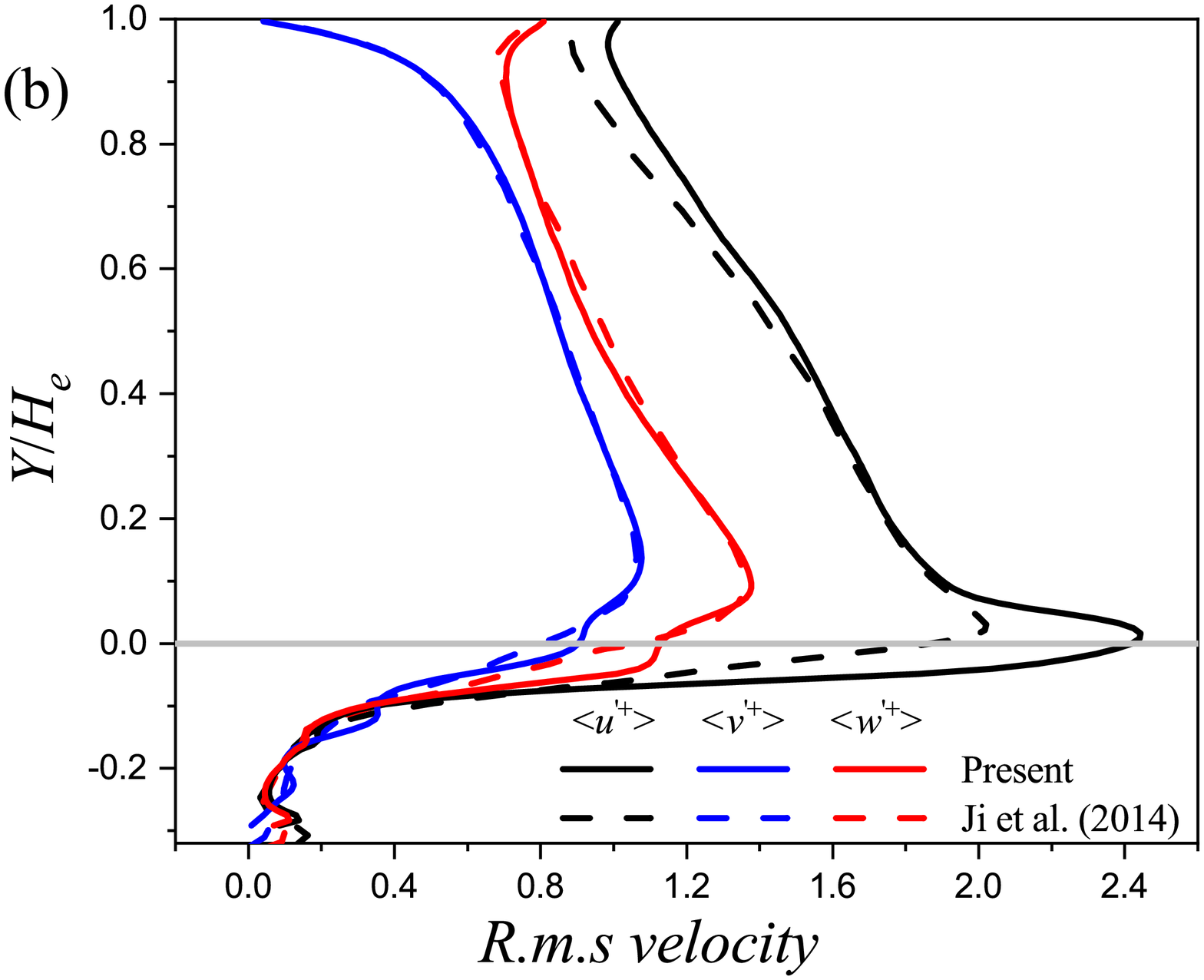}
\hspace{-16mm}
\caption{Profiles of r.m.s. velocities  of the carrier phase: (a) static case and (b) mobile case. The gray line indicates the effective height of the sediment bed.}
\label{fig:rms}
\end{figure}

The root-mean-squared (r.m.s.) velocity profiles of the disperse phase are shown in Fig.~\ref{fig:rms}. The prime superscript indicates the r.m.s. values of fluctuating velocities. The present r.m.s. velocities in the three directions in the static case are in good agreement with those in \cite{ji2014saltation}. Good agreement can also be seen for the mobile case except in the vicinity of the bed surface, where the present r.m.s. values of the carrier phase velocity fluctuations are stronger, especially for the streamwise component. This may be caused by the different particle distributions in the sediment bed or the different collision models noted above.

\begin{figure}[H]
\centering
\includegraphics[width=10cm]{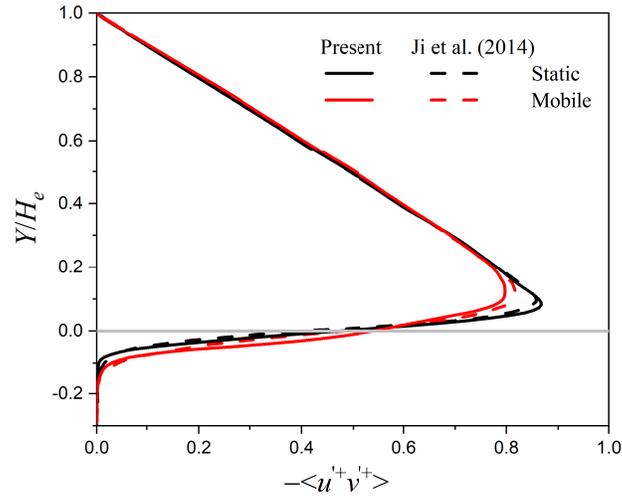}
\caption{Reynolds shear stress profiles of the carrier phase. The gray line indicates the effective height of the sediment bed.}
\label{fig:stress}
\end{figure}

The results of the Reynolds shear stress for the carrier phase are presented in Fig.~\ref{fig:stress}. The present simulation and results in \cite{ji2014saltation} are in good agreement for both the static and mobile cases.

\begin{figure}[H]
\centering
\includegraphics[width=10cm]{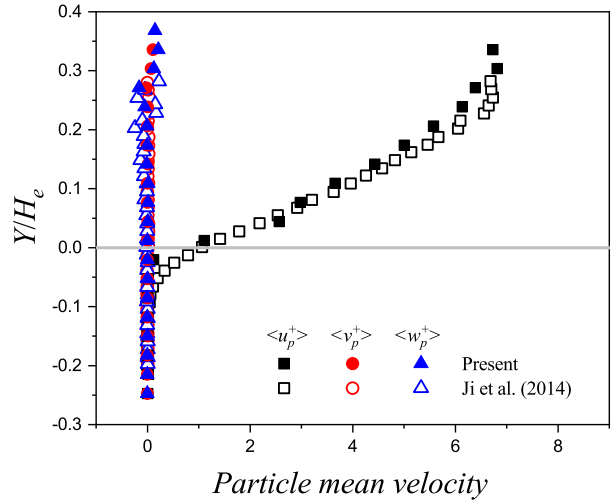}
\caption{Mean velocity profiles of the disperse phase. The gray line indicates the effective height of the sediment bed.}
\label{fig:pu}
\end{figure}

For the disperse phase, we compare the results of the nondimensional transport rate $\phi_p$ and the mean velocity profile. The nondimensional transport rate is defined as $\phi_p =\int_{0}^{H}q(y)dy/\sqrt{(\rho _p/\rho _f-1)gD_p^3}$, in which $q=C \langle u_p \rangle$ is the volume flux density of the disperse phase, with $C$ being the particle volume fraction and $\langle u_p \rangle$ the mean streamwise particle velocity. The present simulation yields $\phi_p=0.0331$, slightly higher than the value of $\phi_p=0.0327$ in \cite{ji2014saltation}. This result is also in good agreement with several empirical bedload transport models and experimental data, with values in the range of 0.01 to 0.04 at ${\it \Theta} =0.065$ \cite[]{wiberg1989model}. Finally, the mean velocity profiles of the disperse phase are shown in Fig.~\ref{fig:pu}. The mean particle velocities in the three directions are in good agreement with the results in \cite{ji2014saltation}.

\subsection{Stochastic features and splash function of saltation particles}
\subsubsection{The extraction method of impact and rebound particles}


\textcolor{black}{After the turbulent flow and sediment transport reach steady state, we continuously record all 4608 particles' information such as vertical height of particles $y_p$, particle velocity $\bm u_p$, particle angular velocity in the spanwise direction $\omega_{p,z}$ in a time period of $15H/u_{\tau}$. Extracting impact and rebound information from these data requires three steps:}

\textcolor{black}{(1) Detection of bed particles and flying particles dynamically since the topography of the mobile sediment bed is changing at every time step. The detection idea is that the particles in contact with the bottom layer (i.e., $y_p<D_p$) of particles directly or indirectly are the bed particles. All remaining particles are the flying particles. The iteration process of detection is as follow,
\begin{align*}
& \text{do \ while}\ N_{bed} \neq N_{temp} \\
& N_{temp}=N_{bed} \\
& N_{bed}=0 \\
& \ \ \text{do} \ i=1,N_p\\
& \ \ \ \ \text{if\ a\ particle\ contact\ with\ any\ one\ of\ the\ bed\ particles,\ then}\\
& \ \ \ \ \ \ \text{it\ become\ a\ bed\ particle}.\\
& \ \ \ \ \ \ N_{bed}=N_{bed}+1\\
& \ \ \ \ \text{else}\\
& \ \ \ \ \ \ \text{it\ is\ the\ flying\ particle}.\\
& \ \ \ \ \text{endif}\\
& \ \ \text{enddo}\\
& \text{enddo}
\end{align*}
where $N_{bed}$ is the number of bed particles, $N_{temp}$ is a temporary variable to record the variation of $N_{bed}$ in the iteration process.}

\textcolor{black}{(2) Distinguishing saltating and rolling particles. The rising distance $h_r$ is often used to distinguish them in experiments \cite[]{wiberg1985theoretical,bohm2006two,auel2017sediment}. If a particle departs from the sediment bed and rises its center exceeds a distance of $h_r$, the particle is assumed to be a saltating particle; otherwise, it is a rolling particle. Following \cite{wiberg1985theoretical}, $h_r/D_p=0.5$ is used here.}

\begin{figure}[htbp]
\centering
   \begin{minipage}{0.49\linewidth}
       \centerline{\includegraphics[width=7.5cm]{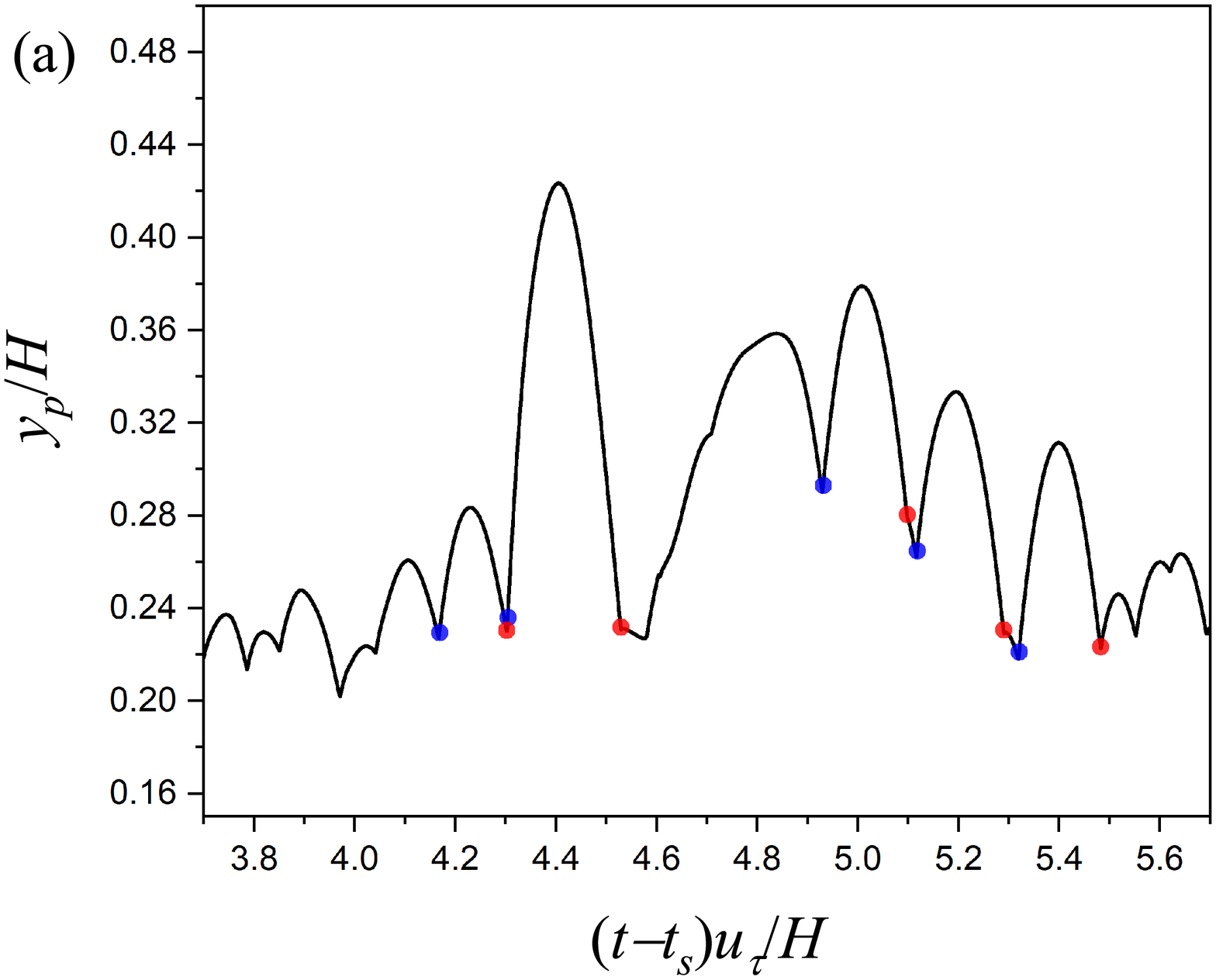}}
       \vspace{-20pt}
       \centerline{\includegraphics[width=7.5cm]{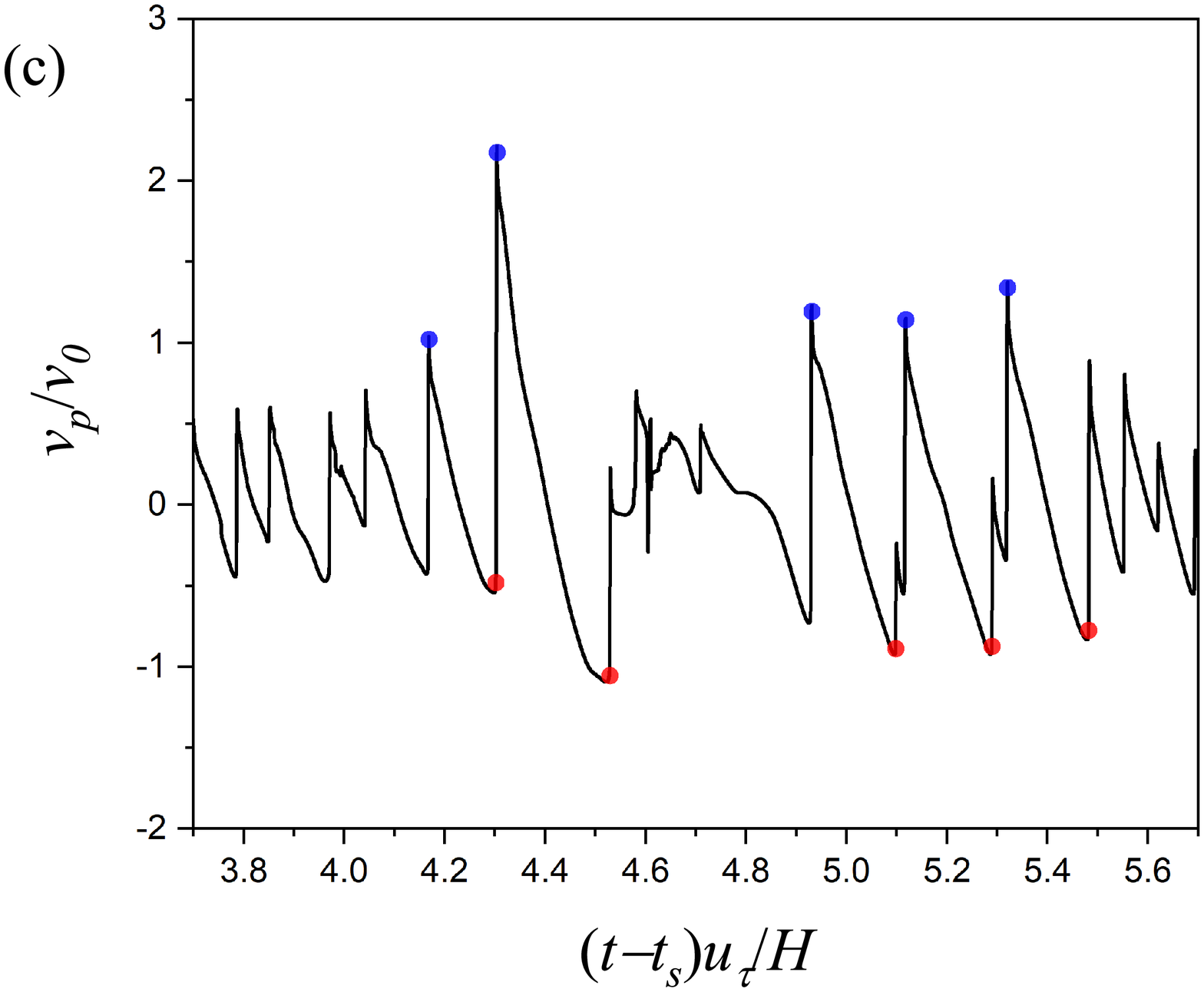}}
   \end{minipage}
      \begin{minipage}{0.49\linewidth}
       \centerline{\includegraphics[width=7.5cm]{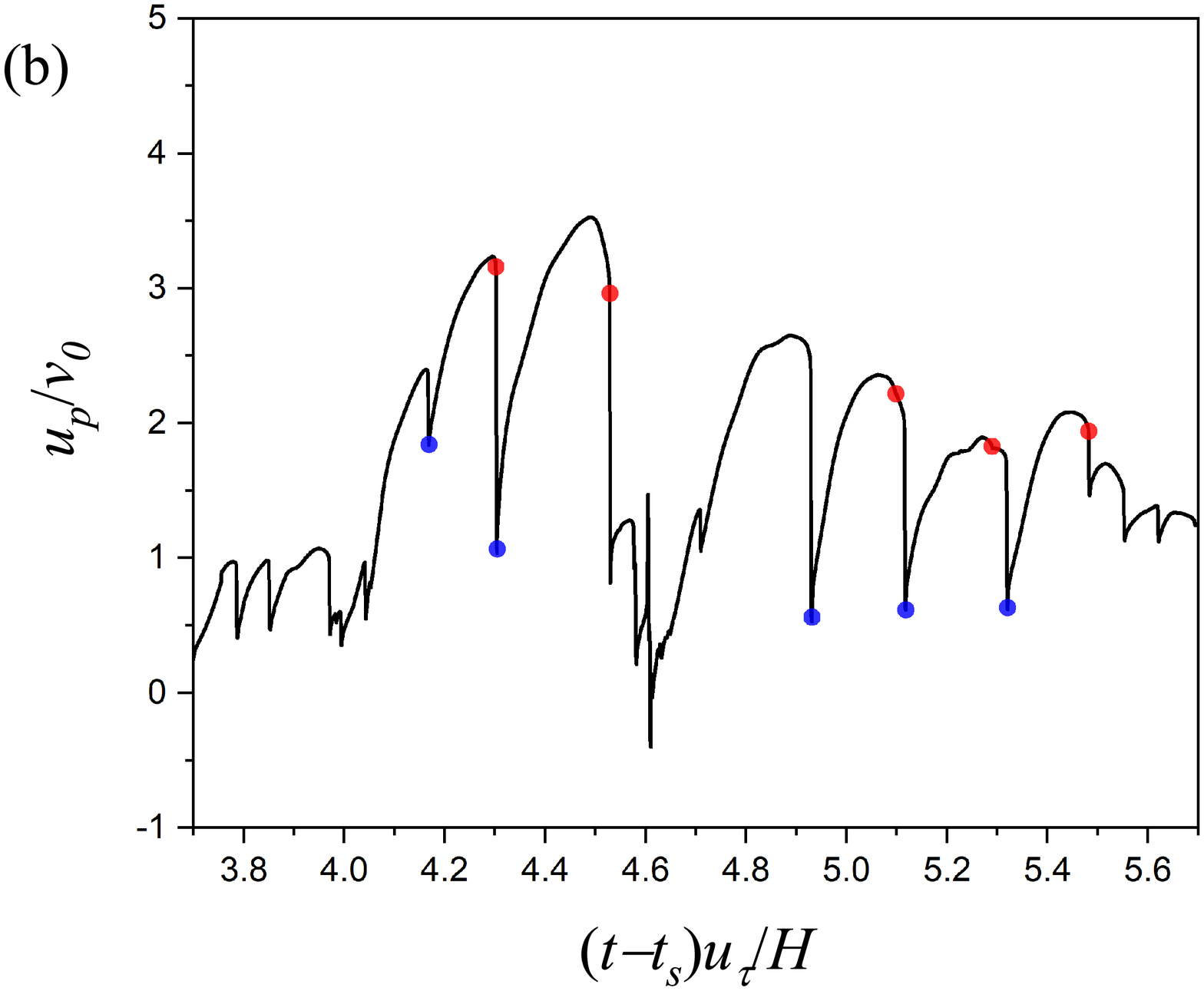}}
       \vspace{-20pt}
       \centerline{\includegraphics[width=7.5cm]{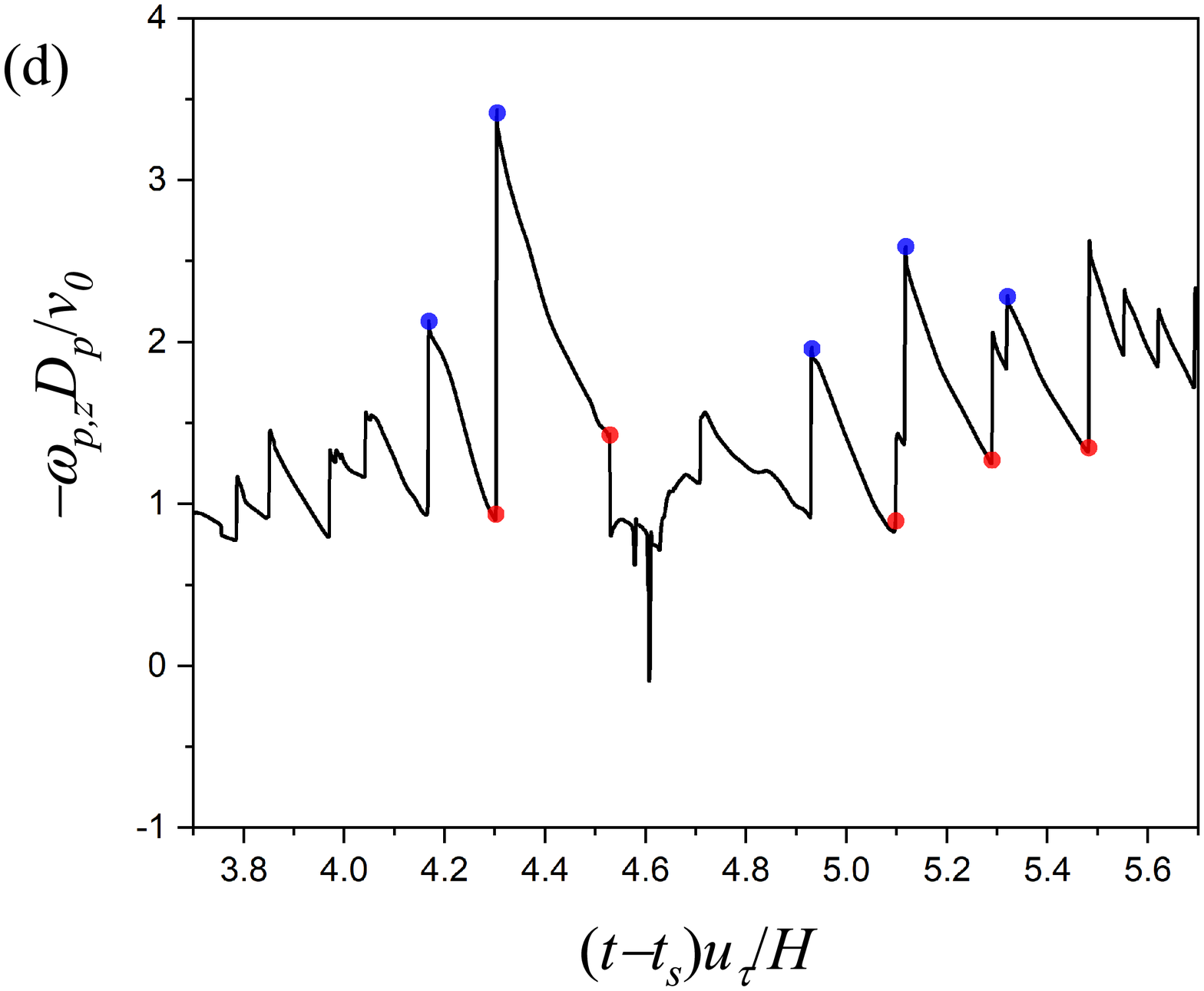}}
   \end{minipage}
   \caption{Partial time evolution of a typical saltating particle: (a) the nondimensional vertical height of the particle, (b) the nondimensional streamwise particle velocity, (c) the nondimensional vertical particle velocity and (d) the nondimensional spanwise angular velocity of the particle. The particle velocity and angular velocity are normalized by $v_0=\sqrt{(1-\rho_f/\rho_p)gD_p}$. $t_s$ is the moment when the turbulent flow and sediment transport reaches steady state. The red points and blue points correspond to the impact moments and departure moments, respectively.}
   \label{fig:extract}
\end{figure}

\textcolor{black}{(3) Extracting impact and rebound information of saltatiing particles. When a saltating particle's state changes from a flying particle to a bed particle, it indicates that this particle impacts the sediment bed. The impact moments correspond to the red points in Fig.~\ref{fig:extract}. On the contrary, when the state changes from a bed particle to a flying particle, this particle departs from the sediment bed. The departure moments correspond to the blue points in Fig.~\ref{fig:extract}. Therefore, one red point and its adjacent blue point to the right constitute a particle-bed impact event of a saltating particle. A criterion time interval $T_{cir}=5T_c$ is used to determine whether the particle is rebounding from the impact event. If the time interval between these two points satisfies $T_i<T_{cir}$, the particle rebounds in the impact event. Otherwise, the particle is enduring contact with the sediment bed and then entrained by the turbulent flow. Since $P \left( T_i<5T_c \left| T_i<15T_c\right. \right) = 82 \% $, $T_{cir}=5T_c$ is a robust criterion. Its variation between $5T_c$ to $15T_c$ has little effect on the statistical features that will be introduced in section 4.3.2.}

\textcolor{black}{Figure~\ref{fig:extract} shows partial time evolution of a typical saltating particle. The translational velocity and angular velocity of particle are normalized by $v_0=\sqrt{(1-\rho_f/\rho_p)gD_p}$. The red points and blue points correspond to the impact moments and departure moments, respectively. These red and blue points accurately capture the impact and departure moments, indicating that the present extraction method is effective and accurate. The dramatic change in the particle velocity and the spanwise angular velocity is due to the collision process between the particle and the sediment bed.}

\textcolor{black}{Through the above method, a total number of 3249 particle-bed impact events is detected in a time period of $15H/u_{\tau}$. These events constitute the samples of the following statistical analysis. The number of samples is much larger compared with 1024 samples in the experiment of \cite{chen2019experimental} and 70 samples in the simulation of \cite{liu2019simulations}, therefore it is sufficient for the following statistical analysis.}

\subsubsection{Stochastic features and splash function of the particle-mobile bed interaction}

\begin{figure}[H]
\centering
\begin{minipage}{0.32\linewidth}
   \centerline{\includegraphics[width=5.cm]{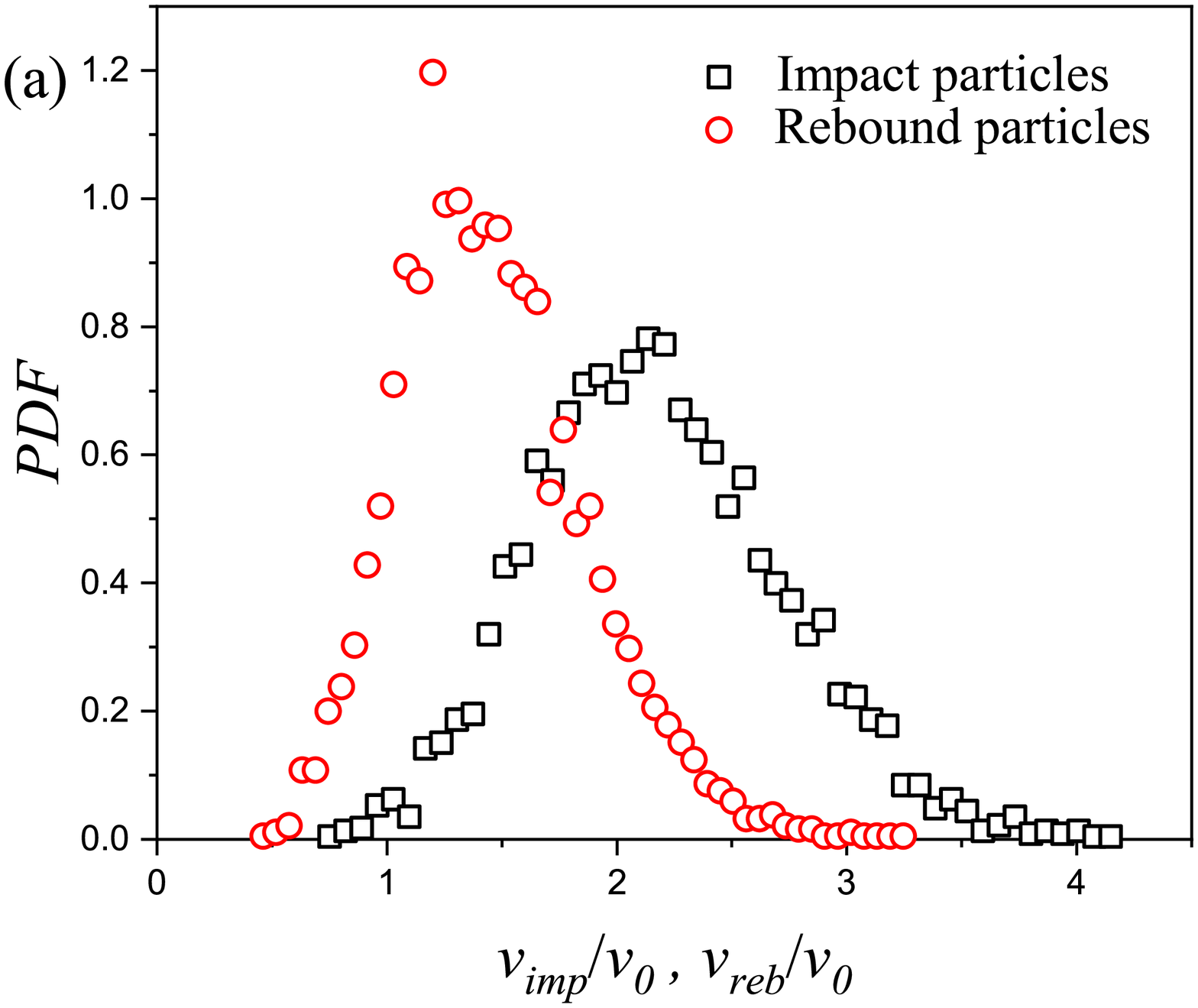}}
\end{minipage}
\begin{minipage}{0.32\linewidth}
   \centerline{\includegraphics[width=5.cm]{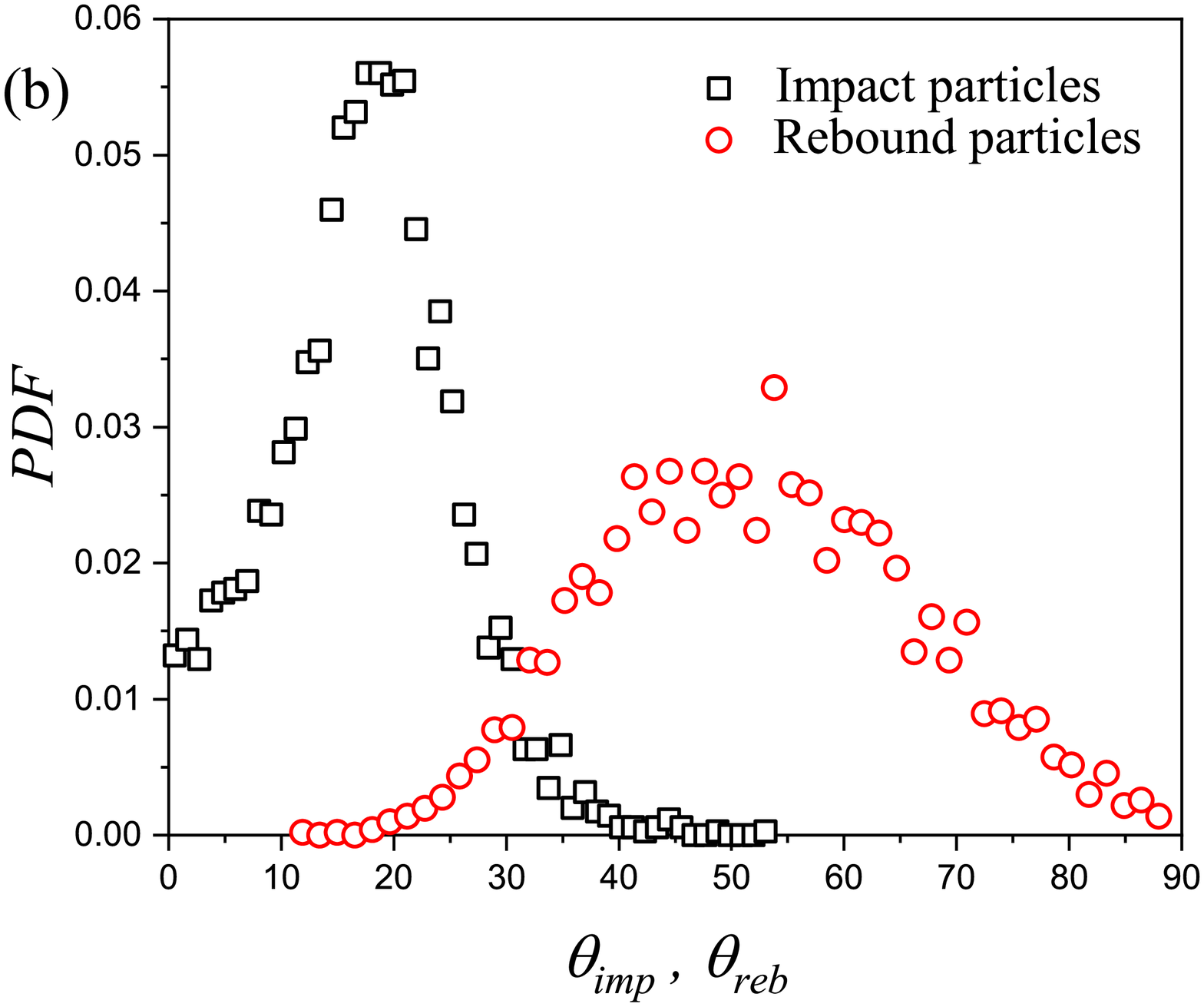}}
\end{minipage}
\begin{minipage}{0.32\linewidth}
   \centerline{\includegraphics[width=5.cm]{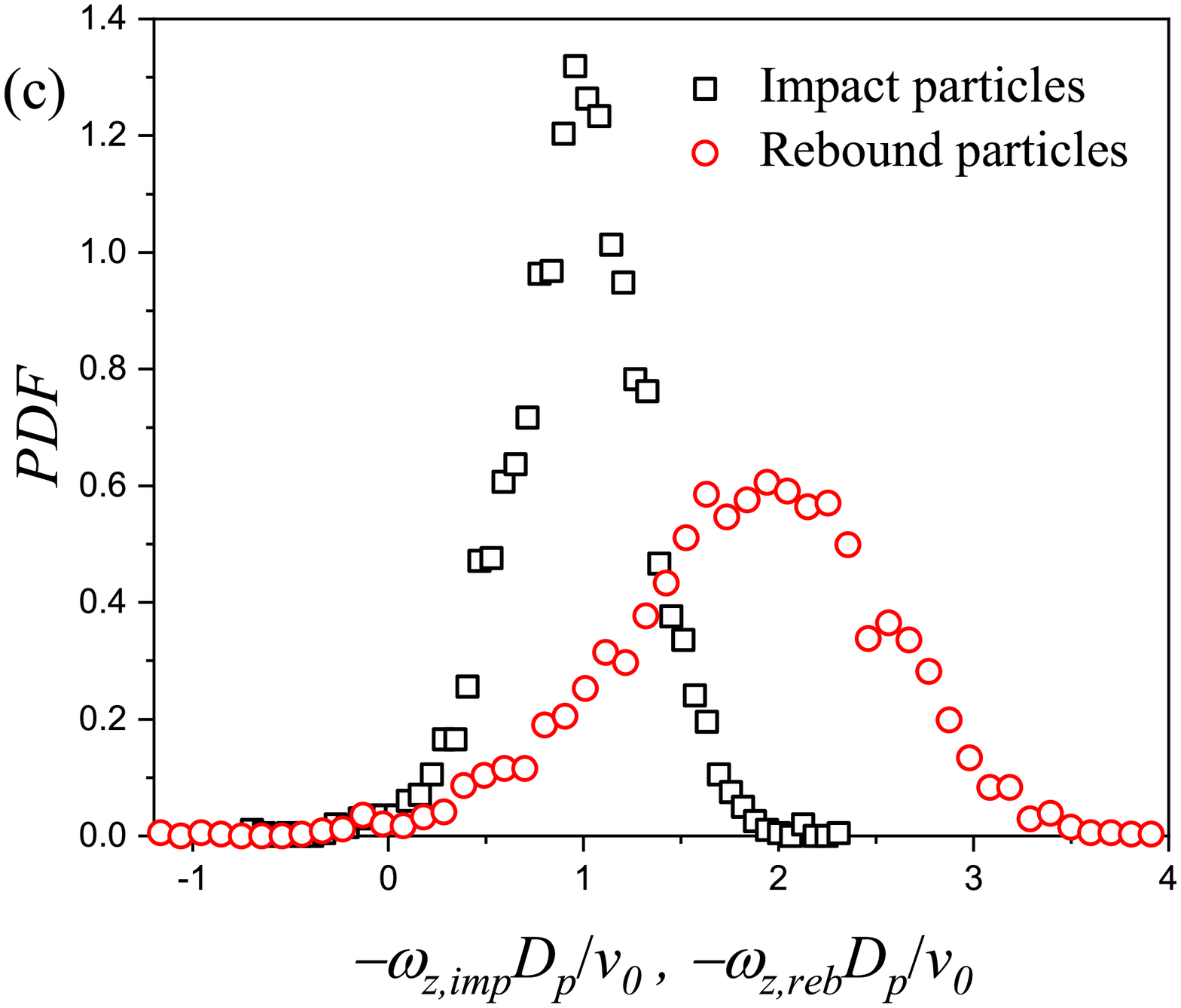}}
\end{minipage}
\caption{Probability density of the nondimensional quantities of the impact and rebound particles: (a) impact and rebound velocity, (b) impact and rebound angle and (c) impact and rebound spanwise angular velocity.}
\label{fig:disir}
\end{figure}

\textcolor{black}{The impact velocity $v_{imp}$ is defined as the magnitude of the particle velocity at the impact moment, and the impact angle $\theta_{imp}$ is defined as the angle between the particle velocity and the horizontal plane at the impact moment. Thus $v_{imp}$ and $\theta_{imp}$ are computed as following,
\begin{equation}
v_{imp}=\sqrt{u_{p,imp,x}^2+u_{p,imp,y}^2+u_{p,imp,z}^2}, 
\end{equation}
\begin{equation}
\theta_{imp}=\arctan(u_{p,imp,y}/\sqrt{u_{p,imp,x}^2+u_{p,imp,z}^2})\times 180^{\circ}/\pi, 
\end{equation}
where the subscript $imp$ indicates the quantities at the impact moment, $u_{p,imp,x}$, $u_{p,imp,y}$, and $u_{p,imp,z}$ are the three components of the particle velocity in the streamwise, vertical and spanwise direction, respectively. The rebound velocity $v_{reb}$ and the rebound angle $\theta_{reb}$ are defined at the rebound moment. In addition to choosing different moments to define $v_{reb}$ and $\theta_{reb}$, the rest are exactly the same as $v_{imp}$ and $\theta_{imp}$. The probability densities of the nondimensional velocity, angle, and spanwise angular velocity for the impact and rebound particles are shown in Fig.~\ref{fig:disir}. {The nondimensional impact velocity $v_{imp}/v_0$, impact angle $\theta_{imp}$, and impact angular velocity in the spanwise direction $-\omega_{z,imp}D_p/v_0$ of all 3249 particle-bed impact events are in the range of $[0.72,4.12]$, $[0.02^{\circ},53.57^{\circ}]$, and $[-0.73,2.34]$, respectively.} To investigate the relationship between the impact velocity (angle, angular velocity) and the rebound velocity (angle, angular velocity), these collision events are equally divided into 10 groups from the minimum value to the maximum value of $v_{imp}/v_0$ ($\theta_{imp}$, $-\omega_{z,imp}D_p/v_0$). To ensure at least 40 samples in each group for analysis, only 5 groups of data that satisfy this condition are used to study the rebound probability and statistical distribution.}

\begin{figure}[H]
\centering
\includegraphics[width=10cm]{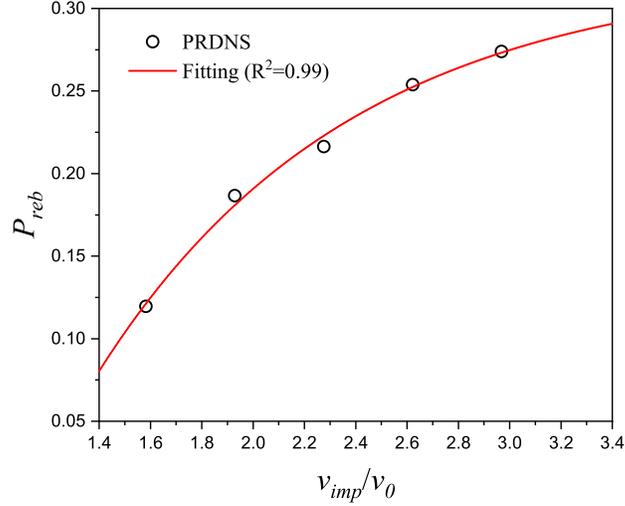}
\caption{Rebound probability of the impact particles. The symbols are the simulation results, and the line is curve fitting result by equation~(14). $R^2$ is the coefficient of determination, which is used to evaluate the fitting results.}
\label{fig:rebp}
\end{figure}

\textcolor{black}{Figure~\ref{fig:rebp} shows the rebound probability $P_{reb}$ of the impact particles. The circles are the simulation results and the red line is curve fitting result by equation~(\ref{eqn:rebp}). There is a negative exponential relationship between $P_{reb}$ and $v_{imp}/v_0$, where $A=1, B=1.01, C=-0.68$. Based on equation~(\ref{eqn:rebp}), it is found that when the nondimensional impact velocity is smaller than $-\ln(1+C/A)/B=1.12$, the impact particles will not rebound from the sediment bed.}

\begin{equation}
\begin{cases} 
P_{reb}=A\left (1-\exp\left (-Bv_{imp}/v_0  \right )   \right )+C ,  & v_{imp}/ v_0>-\ln(1+C/A)/{B}, \\
P_{reb}=0,  & v_{imp}/v_0\leqslant-\ln(1+C/A)/{B},
\end{cases}
\label{eqn:rebp}
\end{equation}

\begin{figure}[H]
\centering
\includegraphics[width=10cm]{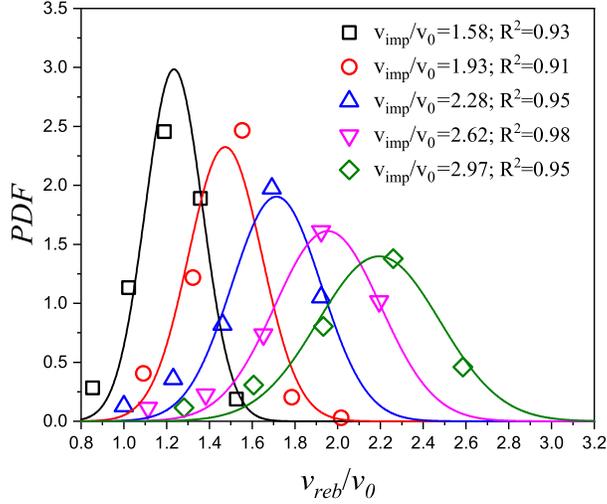}
\caption{Probability density of the nondimensional rebound velocity. The symbols are the simulation results, and the lines are curve fitting results by equation~(15). $R^2$ is the coefficient of determination, which is used to evaluate the fitting results.}
\label{fig:imv_rbv}
\end{figure}

{Figure~\ref{fig:imv_rbv} shows the probability density of the nondimentional rebound velocity for the prescribed values of the impact velocity shown by markers in Fig.~\ref{fig:rebp}}. It obeys the normal distribution and is formulated as equation~(15a). The probability distribution parameters $\mu$ and $\sigma$ vary with $v_{imp}/v_0$. The relationship between them are established as equation~(15b) and equation~(15c).

\begin{align}
prob\left( \frac{v_{reb}}{v_0}\right)&=\frac{1}{\sqrt{2\pi}\sigma}\exp\left(-\frac{(\frac{v_{reb}}{v_0}-\mu)^2}{2\sigma^2}\right) \tag{15a}\\
\mu&=0.6919\frac{v_{imp}}{v_0}+0.1396 \tag{15b}\\
\sigma&=0.1094\frac{v_{imp}}{v_0}-0.0394 \tag{15c}
\label{eqn:imrbv}
\end{align}

\begin{figure}[H]
\centering
\includegraphics[width=10cm]{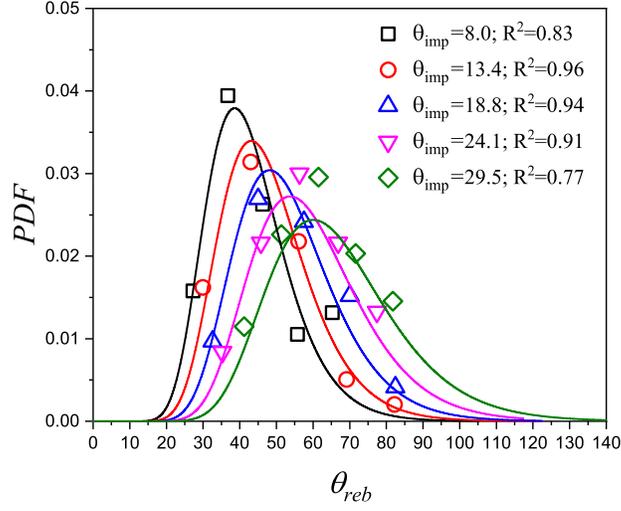}
\caption{Probability density of the rebound angle. The symbols are the simulation results, and the lines are curve fitting results by equation~(16), where $R^2$ is used to evaluate the fitting results.}
\label{fig:ima_rba}
\end{figure}

\textcolor{black}{Figure~\ref{fig:ima_rba} shows the probability density of the rebound angle. It obeys the log-normal distribution and is formulated as equation~(16a). The probability distribution parameter $\mu$ varies with $\theta_{imp}$, and $\sigma$ is a constant. The relationship between them is established as equation~(16b) and equation~(16c).}

\begin{align}
prob\left( \theta_{reb} \right)&=\frac{1}{\sqrt{2\pi}\sigma\theta_{reb}}\exp\left(-\frac{(\ln\theta_{reb}-\mu)^2}{2\sigma^2}\right) \tag{16a}\\
\mu&=0.0206\theta_{imp}+3.5571 \tag{16b}\\
\sigma&=0.2632 \tag{16c}
\label{eqn:imrba}
\end{align}

\begin{figure}[H]
\centering
\includegraphics[width=10cm]{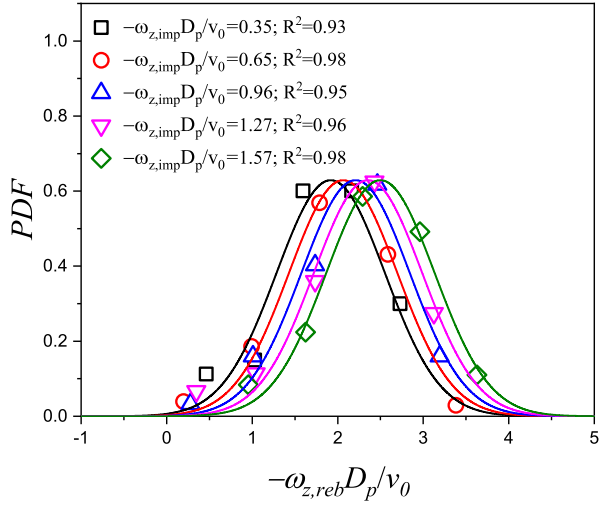}
\caption{Probability density of the spanwise nondimensional rebound angular velocity. The symbols are the simulation results, and the lines are curve fitting results by equation~(17), where $R^2$ is used to evaluate the fitting results.}
\label{fig:imav_rbav}
\end{figure}

\textcolor{black}{Figure~\ref{fig:imav_rbav} shows the probability density of the spanwise nondimensional rebound angular velocity. It obeys the normal distribution and is formulated as equation~(17a). The probability distribution parameter $\mu$ varies with $-\omega_{z,imp}D_p/v_0$, and $\sigma$ is a constant. The relationship between them is established as equation~(17b) and equation~(17c).}

\begin{align}
prob\left( \frac{-\omega_{z,reb}D_p}{v_0}\right)&=\frac{1}{\sqrt{2\pi}\sigma}\exp\left(-\frac{(\frac{-\omega_{z,reb}D_p}{v_0}-\mu)^2}{2\sigma^2}\right) \tag{17a}\\
\mu&=0.4736\left( \frac{-\omega_{z,imp}D_p}{v_0}\right)+1.7547 \tag{17b}\\
\sigma&=0.6342 \tag{17c}
\label{eqn:imrbav}
\end{align}

\section{Conclusions}
{In the present study, the parallel approach of \cite{darmana2006parallelization} originally developed for point particle simulations is modified and implemented into particle resolved simulations. The hybrid parallel approach improves the load imbalance of sediment transport where they are concentrated at the bottom of the fluid domain.} The memory requirement of the hybrid approach is reduced by a novel memory optimization technique for spherical particles with equal size. The scaling analysis for parallel performance is conducted for a challenging sediment transport case with more than a million spherical particles. The results show good performance of the hybrid parallel approach. The scaling of the overall and the most time-consuming subprocedures such as Poisson solver, IB force, and DEM demonstrate good scalability of the code. The present hybrid approach is applied to several benchmark cases to ascertain the accuracy. The results show good agreement with the experimental and simulation data in the literature.

\textcolor{black}{Furthermore, a turbulent flow over an erodible sediment bed is simulated. An extraction method is proposed to distinguish the saltating and rolling particles and extract impact and rebound information of the particle-mobile bed interaction. It can accurately capture the moments of impact and rebound. The probability distribution functions (PDF) of several important saltation parameters such as velocity, angle, and spanwise angular velocity for the impact and rebound events are presented. The splash functions are established for the particle-mobile bed interaction in the turbulent flow, which was rarely investigated before. The rebound probability has a negative exponential relationship with impact velocity. The critical impact velocity is determined by the curve fitting. If the impact velocity is smaller than the critical value, the impact particles will not rebound from the sediment bed. The rebound velocity follows a normal distribution with different impact velocity, the rebound angle follows a lognormal distribution with different impact angle, and the spanwise rebound angular velocity follows a normal distribution with different spanwise impact angular velocity. Quantitative characterization of the distribution parameters as a function of the impact information is also given. The splash functions are helpful to model the complex particle-bed interactions in the turbulent flow.}

\section{Acknowledgments}
This work was financially supported by grants from the National Natural Science Foundation of China (Nos. 92052202 and 11972175), the National Numerical Windtunnel project and the Fundamental Research Funds for the Central Universities (lzujbky-2020-17). The computation was performed at the Tianhe-2A supercomputer of the National Supercomputer Center in Guangzhou.

\section*{Appendix A. Definitions for particle-particle and particle-wall collisions}

The quantities used in particle collisions are defined below. Some definitions depend on whether the interaction is between particle $p$ and the wall (P-W) or between particle $p$ and particle $q$ (P-P). They are as follows:

$\bm{n}$ - normal unit vector of contact
\begin{alignat}{2}
&\bm n=\frac{\bm x_q-\bm x_p}{\left | \bm x_q-\bm x_p \right | },  \quad &&\text{(P-P)}, \tag{A.1}\\
&\bm n=\frac{\bm x_w-\bm x_p}{\left | \bm x_w-\bm x_p \right | }, &&\text{(P-W)}, \tag{A.2}
\end{alignat}

$\delta_n$ - distance between two surfaces
\begin{alignat}{2}
&\delta _{n}=\left | \bm x_q- \bm x_p \right |-R_p-R_q,  \quad &&\text{(P-P)}, \tag{A.3}\\
&\delta _{n}=\left | \bm x_w- \bm x_p \right |-R_p, &&\text{(P-W)}, \tag{A.4}
\end{alignat}

$\bm u_{cp}$ - relative velocity of the contact point
\begin{alignat}{2}
&\bm u_{cp}=\bm u_p-\bm u_q+ R_{p} \bm \omega_p \times \bm n +R_{q} \bm \omega_q \times \bm n, \quad &&\text{(P-P)}, \tag{A.5}\\
&\bm u_{cp}=\bm u_p+R_{p} \bm \omega_p \times \bm n, &&\text{(P-W)}, \tag{A.6}
\end{alignat}

$\bm u_{cp,n}$ - normal component of $\bm u_{cp}$
\begin{alignat}{1}
&\bm u_{cp,n}=(\bm u_{cp} \cdot \bm n) \bm n, \tag{A.7}
\end{alignat}

$\bm u_{cp,t}$ - tangential component of $\bm u_{cp}$
\begin{alignat}{1}
&\bm u_{cp,t}=\bm u_{cp}-\bm u_{cp,n}, \tag{A.8}
\end{alignat}
and $\bm \delta_t$ - tangential displacement of the contact point.

The direction of the tangential unit vector changes at different time steps. Therefore, we need to rotate the displacement from the previous time step onto a plane tangent to $\bm n$. $\bm \delta_t$ is calculated following \cite{biegert2017collision} as
\begin{alignat}{3}
&\widetilde{\bm \delta _{t}} =\bm \delta _{t}^{k-1}-(\bm \delta _{t}^{k-1} \cdot \bm n)\bm n, \tag{A.9}\\
&\widehat{\bm \delta _{t}} =\frac{\left |\bm \delta _{t}^{k-1} \right | }{\left |\widetilde{\bm \delta _{t}} \right |} \widetilde{\bm \delta _{t}},  \tag{A.10}\\
&\bm \delta _{t}^{k}=\widehat{\bm \delta _{t}}+\Delta t \bm u_{cp,t}. \tag{A.11}
\end{alignat}

\section*{Appendix B. Averaging operations for flow and particle variables}

\subsection*{B.1. Averaging operations for flow variables}
Before averaging the flow variables, an indicator function needs to be defined as $\phi _f(\bm x,t)$ to distinguish the Eulerian grid point at a position $\bm x$ that is inside or outside of a particle, following \cite{kidanemariam2014interface}, as 
\begin{equation}
\phi _f(\bm x,t)= \begin{cases} 
1,  & \text{ if } \bm x \ \text{is outside a particle,} \\
0,  & \text{ otherwise.} \tag{B.1}
\end{cases}
\end{equation}
Based on the indicator function $\phi _f(\bm x,t)$, only the flow data outside of the particle are accounted for as follows:
\begin{equation}
n_f(y_j)=\sum \limits_{n=1}^{N_t}\sum \limits_{i=1}^{N_x}\sum \limits_{k=1}^{N_z}\phi _f(\bm x_{ijk},t^n), \tag{B.2}
\end{equation}
where $n_f(y_j)$ is the total grid number in the $x-z$ plane over $N_t$ time steps for flow statistics at a given height $y_j$. Therefore, the ensemble average of the flow variables $\xi_f(\bm x,t)$ can be defined as
\begin{equation}
\left \langle\xi_f\right \rangle (y_j)=\frac{1}{n_f(y_j)}\sum \limits_{n=1}^{N_t}\sum \limits_{i=1}^{N_x}\sum \limits_{k=1}^{N_z} \xi_f(\bm x_{ijk},t^n)\phi _f(\bm x_{ijk},t^n), \tag{B.3}
\end{equation}
where the operator $\left \langle \cdot \right \rangle$ indicates the average over the $x-z$ plane and time.

\subsection*{B.2. Averaging operations for particle variables}
Particle variables are averaged over the particle number within each bin. The bin is generated by dividing $H$ by the thickness $\Delta h$. An indicator function $\phi _{p}^j(y_p,t)$ is defined to distinguish the center height $y_p$ of a particle inside or outside of a bin with index $j$ as follows:
\begin{equation}
\phi_{p}^j(y_p,t)= \begin{cases} 
1,  & \text{ if } (j-1)\Delta h \leqslant y<j\Delta h \\
0,  & \text{ otherwise.} \tag{B.4}
\end{cases}
\end{equation}
Based on the indicator function $\phi _{p}^j(y_p,t)$, the particle number in each bin can be calculated as
\begin{equation}
n_p^j=\sum \limits_{n=1}^{N_t}\sum \limits_{l=1}^{N_p}\phi _{p}^j(y_p^l,t^n), \tag{B.5}
\end{equation}
where $n_p^j$ is the total particle number in bin $j$ over $N_t$ time steps. Therefore, the averaged particle variable $\xi_p$ can be defined as
\begin{equation}
\left \langle\xi_p(y^j)\right \rangle=\frac{1}{n_p^j}\sum \limits_{n=1}^{N_t}\sum \limits_{l=1}^{N_p}\phi _{p}^j(y_p^l,t^n)\xi_p^l(t^n). \tag{B.6}
\end{equation}
A bin thickness of $\Delta h=D_p/4$ is chosen here. If $n_p^j/N_t<1$, the $\left \langle\xi_p(y^j)\right \rangle$ calculated in bin $j$ is not shown owing to insufficient particle samples.


\bibliographystyle{elsarticle-harv}
\bibliography{ref}

\begin{thebibliography}{99}
\expandafter\ifx\csname natexlab\endcsname\relax\def\natexlab#1{#1}\fi
\expandafter\ifx\csname url\endcsname\relax
  \def\url#1{\texttt{#1}}\fi
\expandafter\ifx\csname urlprefix\endcsname\relax\def\urlprefix{URL }\fi

\bibitem[{Akiki and Balachandar(2016)}]{akiki2016immersed}
Akiki, G., Balachandar, S., 2016. Immersed boundary method with non-uniform
  distribution of {L}agrangian markers for a non-uniform {E}ulerian mesh.
  Journal of Computational Physics 307, 34--59.

\bibitem[{Ammi et~al.(2009)Ammi, Oger, Beladjine, and Valance}]{ammi2009three}
Ammi, M., Oger, L., Beladjine, D., Valance, A., 2009. Three-dimensional
  analysis of the collision process of a bead on a granular packing. Physical
  Review E 79~(2), 021305.

\bibitem[{Amritkar et~al.(2014)Amritkar, Deb, and
  Tafti}]{amritkar2014efficient}
Amritkar, A., Deb, S., Tafti, D., 2014. Efficient parallel cfd-dem simulations
  using openmp. Journal of Computational Physics 256, 501--519.

\bibitem[{Anderson and Haff(1991)}]{anderson1991wind}
Anderson, R.~S., Haff, P., 1991. Wind modification and bed response during
  saltation of sand in air. In: Aeolian Grain Transport 1. Springer, pp.
  21--51.

\bibitem[{Anderson and Haff(1988)}]{anderson1988simulation}
Anderson, R.~S., Haff, P.~K., 1988. Simulation of eolian saltation. Science
  241~(4867), 820--823.

\bibitem[{Auel et~al.(2017)Auel, Albayrak, Sumi, and Boes}]{auel2017sediment}
Auel, C., Albayrak, I., Sumi, T., Boes, R.~M., 2017. Sediment transport in
  high-speed flows over a fixed bed: 1. particle dynamics. Earth Surface
  Processes and Landforms 42~(9), 1365--1383.

\bibitem[{Bagnold(1941)}]{bagnold1941physics}
Bagnold, R.~A., 1941. The physics of blown sand and desert dunes. William
  Morrow \& Company, New York.

\bibitem[{Berk and Coletti(2020)}]{berk2020transport}
Berk, T., Coletti, F., 2020. Transport of inertial particles in
  high-{R}eynolds-number turbulent boundary layers. Journal of Fluid Mechanics
  903, A18.

\bibitem[{Berzi et~al.(2016)Berzi, Jenkins, and Valance}]{berzi2016periodic}
Berzi, D., Jenkins, J.~T., Valance, A., 2016. Periodic saltation over
  hydrodynamically rough beds: Aeolian to aquatic. Journal of Fluid Mechanics
  786, 190--209.

\bibitem[{Biegert et~al.(2017)Biegert, Vowinckel, and
  Meiburg}]{biegert2017collision}
Biegert, E., Vowinckel, B., Meiburg, E., 2017. A collision model for
  grain-resolving simulations of flows over dense, mobile, polydisperse
  granular sediment beds. Journal of Computational Physics 340, 105--127.

\bibitem[{Bo et~al.(2017)Bo, Fu, Liu, and Zheng}]{bo2017improved}
Bo, T., Fu, L., Liu, L., Zheng, X., 2017. An improved numerical model suggests
  potential differences of wind-blown sand between on {E}arth and {M}ars.
  Journal of Geophysical Research: Atmospheres 122~(11), 5823--5836.

\bibitem[{B{\"o}hm et~al.(2006)B{\"o}hm, Frey, Ducottet, Ancey, Jodeau, and
  Reboud}]{bohm2006two}
B{\"o}hm, T., Frey, P., Ducottet, C., Ancey, C., Jodeau, M., Reboud, J.-L.,
  2006. Two-dimensional motion of a set of particles in a free surface flow
  with image processing. Experiments in fluids 41~(1), 1--11.

\bibitem[{Bragg et~al.(2021)Bragg, Richter, and Wang}]{bragg2021mechanisms}
Bragg, A.~D., Richter, D.~H., Wang, G., 2021. Mechanisms governing the settling
  velocities and spatial distributions of inertial particles in wall-bounded
  turbulence. Physical Review Fluids 6~(6), 064302.

\bibitem[{Breugem(2012)}]{breugem2012second}
Breugem, W.-P., 2012. A second-order accurate immersed boundary method for
  fully resolved simulations of particle-laden flows. Journal of Computational
  Physics 231~(13), 4469--4498.

\bibitem[{Capecelatro and Desjardins(2013)}]{capecelatro2013euler}
Capecelatro, J., Desjardins, O., 2013. An euler--lagrange strategy for
  simulating particle-laden flows. Journal of Computational Physics 238, 1--31.

\bibitem[{Chen et~al.(2019)Chen, Zhang, Huang, and Xu}]{chen2019experimental}
Chen, Y., Zhang, J., Huang, N., Xu, B., 2019. An experimental study on splash
  functions of natural sand-bed collision. Journal of Geophysical Research:
  Atmospheres 124~(13), 7226--7235.

\bibitem[{Chien and Wan(1999)}]{chien1999mechanics}
Chien, N., Wan, Z., 1999. Mechanics of sediment transport. ASCE Press, Reston.

\bibitem[{Chung et~al.(2021)Chung, Hutchins, Schultz, and
  Flack}]{chung2021predicting}
Chung, D., Hutchins, N., Schultz, M.~P., Flack, K.~A., 2021. Predicting the
  drag of rough surfaces. Annual Review of Fluid Mechanics 53, 439--471.

\bibitem[{Costa et~al.(2015)Costa, Boersma, Westerweel, and
  Breugem}]{costa2015collision}
Costa, P., Boersma, B.~J., Westerweel, J., Breugem, W.-P., 2015. Collision
  model for fully resolved simulations of flows laden with finite-size
  particles. Physical Review E 92~(5), 053012.

\bibitem[{Costa et~al.(2018)Costa, Picano, Brandt, and
  Breugem}]{costa2018effects}
Costa, P., Picano, F., Brandt, L., Breugem, W.-P., 2018. Effects of the finite
  particle size in turbulent wall-bounded flows of dense suspensions. Journal
  of Fluid Mechanics 843, 450--478.

\bibitem[{Cui et~al.(2018)Cui, Yang, Jiang, Huang, and Shen}]{cui2018sharp}
Cui, Z., Yang, Z., Jiang, H.-Z., Huang, W.-X., Shen, L., 2018. A
  sharp-interface immersed boundary method for simulating incompressible flows
  with arbitrarily deforming smooth boundaries. International Journal of
  Computational Methods 15~(1), 1750080.

\bibitem[{Dai et~al.(2019)Dai, Reimann, Hanaor, Ferrero, and
  Gan}]{dai2019modes}
Dai, W., Reimann, J., Hanaor, D., Ferrero, C., Gan, Y., 2019. Modes of wall
  induced granular crystallisation in vibrational packing. Granular Matter
  21~(2), 1--16.

\bibitem[{Darmana et~al.(2006)Darmana, Deen, and
  Kuipers}]{darmana2006parallelization}
Darmana, D., Deen, N.~G., Kuipers, J., 2006. Parallelization of an
  {E}uler--{L}agrange model using mixed domain decomposition and a mirror
  domain technique: {A}pplication to dispersed gas--liquid two-phase flow.
  Journal of Computational Physics 220~(1), 216--248.

\bibitem[{Deen et~al.(2007)Deen, Annaland, Van~der Hoef, and
  Kuipers}]{deen2007review}
Deen, N., Annaland, M. V.~S., Van~der Hoef, M.~A., Kuipers, J., 2007. Review of
  discrete particle modeling of fluidized beds. Chemical engineering science
  62~(1-2), 28--44.

\bibitem[{Dufresne et~al.(2020)Dufresne, Moureau, Lartigue, and
  Simonin}]{dufresne2020massively}
Dufresne, Y., Moureau, V., Lartigue, G., Simonin, O., 2020. A massively
  parallel {CFD/DEM} approach for reactive gas-solid flows in complex
  geometries using unstructured meshes. Computers \& Fluids 198, 104402.

\bibitem[{Dur{\'a}n et~al.(2012)Dur{\'a}n, Andreotti, and
  Claudin}]{duran2012numerical}
Dur{\'a}n, O., Andreotti, B., Claudin, P., 2012. Numerical simulation of
  turbulent sediment transport, from bed load to saltation. Physics of Fluids
  24~(10), 103306.

\bibitem[{Dur{\'a}n et~al.(2011)Dur{\'a}n, Claudin, and
  Andreotti}]{duran2011aeolian}
Dur{\'a}n, O., Claudin, P., Andreotti, B., 2011. On aeolian transport:
  {G}rain-scale interactions, dynamical mechanisms and scaling laws. Aeolian
  Research 3~(3), 243--270.

\bibitem[{Finn et~al.(2016)Finn, Li, and Apte}]{finn2016particle}
Finn, J.~R., Li, M., Apte, S.~V., 2016. Particle based modelling and simulation
  of natural sand dynamics in the wave bottom boundary layer. Journal of Fluid
  Mechanics 796, 340--385.

\bibitem[{Gondret et~al.(2002)Gondret, Lance, and Petit}]{gondret2002bouncing}
Gondret, P., Lance, M., Petit, L., 2002. Bouncing motion of spherical particles
  in fluids. Physics of Fluids 14~(2), 643--652.

\bibitem[{Gopalakrishnan and Tafti(2013)}]{gopalakrishnan2013development}
Gopalakrishnan, P., Tafti, D., 2013. Development of parallel {DEM} for the open
  source code {MFIX}. Powder Technology 235, 33--41.

\bibitem[{Graf(1984)}]{graf1984hydraulics}
Graf, W.~H., 1984. Hydraulics of sediment transport. Water Resources
  Publication, Colorado.

\bibitem[{Ho et~al.(2011)Ho, Valance, Dupont, and El~Moctar}]{ho2011scaling}
Ho, T.~D., Valance, A., Dupont, P., El~Moctar, A.~O., 2011. Scaling laws in
  aeolian sand transport. Physical Review Letters 106~(9), 094501.

\bibitem[{Huang and Zheng(2003)}]{huang2003effects}
Huang, N., Zheng, X.-J., 2003. Effects of wind-blown sand movement and magnus
  force on effective roughness. Journal of Desert Research 23~(6), 616.

\bibitem[{Jain et~al.(2021)Jain, Tschisgale, and Fr{\"o}hlich}]{jain2021impact}
Jain, R., Tschisgale, S., Fr{\"o}hlich, J., 2021. Impact of shape: {DNS} of
  sediment transport with non-spherical particles. Journal of Fluid Mechanics
  916, A38.

\bibitem[{Ji et~al.(2014)Ji, Munjiza, Avital, Xu, and
  Williams}]{ji2014saltation}
Ji, C., Munjiza, A., Avital, E., Xu, D., Williams, J., 2014. Saltation of
  particles in turbulent channel flow. Physical Review E 89~(5), 052202.

\bibitem[{Jim{\'e}nez(2004)}]{jimenez2004turbulent}
Jim{\'e}nez, J., 2004. Turbulent flows over rough walls. Annual Review of Fluid
  Mechanics 36, 173--196.

\bibitem[{Joseph and Hunt(2004)}]{joseph2004oblique}
Joseph, G., Hunt, M., 2004. Oblique particle-wall collisions in a liquid.
  Journal of Fluid Mechanics 510, 71--93.

\bibitem[{Kadivar et~al.(2021)Kadivar, Tormey, and
  McGranaghan}]{kadivar2021review}
Kadivar, M., Tormey, D., McGranaghan, G., 2021. A review on turbulent flow over
  rough surfaces: {F}undamentals and theories. International Journal of
  Thermofluids 10, 100077.

\bibitem[{Kafui et~al.(2011)Kafui, Johnson, Thornton, and
  Seville}]{kafui2011parallelization}
Kafui, D., Johnson, S., Thornton, C., Seville, J., 2011. Parallelization of a
  {L}agrangian--{E}ulerian {DEM/CFD} code for application to fluidized beds.
  Powder Technology 207~(1-3), 270--278.

\bibitem[{Kempe and Fr{\"o}hlich(2012{\natexlab{a}})}]{kempe2012collision}
Kempe, T., Fr{\"o}hlich, J., 2012{\natexlab{a}}. Collision modelling for the
  interface-resolved simulation of spherical particles in viscous fluids.
  Journal of Fluid Mechanics 709, 445--489.

\bibitem[{Kempe and Fr{\"o}hlich(2012{\natexlab{b}})}]{kempe2012improved}
Kempe, T., Fr{\"o}hlich, J., 2012{\natexlab{b}}. An improved immersed boundary
  method with direct forcing for the simulation of particle laden flows.
  Journal of Computational Physics 231~(9), 3663--3684.

\bibitem[{Kidanemariam and
  Uhlmann(2014{\natexlab{a}})}]{kidanemariam2014direct}
Kidanemariam, A.~G., Uhlmann, M., 2014{\natexlab{a}}. Direct numerical
  simulation of pattern formation in subaqueous sediment. Journal of Fluid
  Mechanics 750, R2.

\bibitem[{Kidanemariam and
  Uhlmann(2014{\natexlab{b}})}]{kidanemariam2014interface}
Kidanemariam, A.~G., Uhlmann, M., 2014{\natexlab{b}}. Interface-resolved direct
  numerical simulation of the erosion of a sediment bed sheared by laminar
  channel flow. International Journal of Multiphase Flow 67, 174--188.

\bibitem[{Kidanemariam and Uhlmann(2017)}]{kidanemariam2017formation}
Kidanemariam, A.~G., Uhlmann, M., 2017. Formation of sediment patterns in
  channel flow: minimal unstable systems and their temporal evolution. Journal
  of Fluid Mechanics 818, 716--743.

\bibitem[{Kim and Moin(1985)}]{kim1985application}
Kim, J., Moin, P., 1985. Application of a fractional-step method to
  incompressible {N}avier-{S}tokes equations. Journal of Computational Physics
  59~(2), 308--323.

\bibitem[{Kok et~al.(2012)Kok, Parteli, Michaels, and Karam}]{kok2012physics}
Kok, J.~F., Parteli, E.~J., Michaels, T.~I., Karam, D.~B., 2012. The physics of
  wind-blown sand and dust. Reports on Progress in Physics 75~(10), 106901.

\bibitem[{Kok and Renno(2009)}]{kok2009comprehensive}
Kok, J.~F., Renno, N.~O., 2009. A comprehensive numerical model of steady state
  saltation (comsalt). Journal of Geophysical Research: Atmospheres 114~(D17).

\bibitem[{Lanigan et~al.(2016)Lanigan, Stout, and
  Anderson}]{lanigan2016atmospheric}
Lanigan, D., Stout, J., Anderson, W., 2016. Atmospheric stability and diurnal
  patterns of aeolian saltation on the {L}lano {E}stacado. Aeolian Research 21,
  131--137.

\bibitem[{Le~Roux and Rojas(2007)}]{le2007sediment}
Le~Roux, J.~P., Rojas, E.~M., 2007. Sediment transport patterns determined from
  grain size parameters: {O}verview and state of the art. Sedimentary Geology
  202~(3), 473--488.

\bibitem[{Lee et~al.(2006)Lee, Lin, Yunyou, and Wenwang}]{lee2006three}
Lee, H.-Y., Lin, Y.-T., Yunyou, J., Wenwang, H., 2006. On three-dimensional
  continuous saltating process of sediment particles near the channel bed.
  Journal of Hydraulic Research 44~(3), 374--389.

\bibitem[{Liu et~al.(2019)Liu, Liu, and Fu}]{liu2019simulations}
Liu, D., Liu, X., Fu, X., 2019. Les-dem simulations of sediment saltation in a
  rough-wall turbulent boundary layer. Journal of Hydraulic Research 57~(6),
  786--797.

\bibitem[{Luo et~al.(2007)Luo, Wang, Fan, and Cen}]{luo2007full}
Luo, K., Wang, Z., Fan, J., Cen, K., 2007. Full-scale solutions to
  particle-laden flows: Multidirect forcing and immersed boundary method.
  Physical Review E 76~(6), 066709.

\bibitem[{Merritt et~al.(2003)Merritt, Letcher, and
  Jakeman}]{merritt2003review}
Merritt, W.~S., Letcher, R.~A., Jakeman, A.~J., 2003. A review of erosion and
  sediment transport models. Environmental Modelling \& Software 18~(8-9),
  761--799.

\bibitem[{Mitha et~al.(1986)Mitha, Tran, Werner, and Haff}]{mitha1986grain}
Mitha, S., Tran, M., Werner, B., Haff, P., 1986. The grain-bed impact process
  in aeolian saltation. Acta Mechanica 63~(1), 267--278.

\bibitem[{Mordant and Pinton(2000)}]{mordant2000velocity}
Mordant, N., Pinton, J.-F., 2000. Velocity measurement of a settling sphere.
  The European Physical Journal B-Condensed Matter and Complex Systems 18~(2),
  343--352.

\bibitem[{Munjiza and Andrews(2000)}]{munjiza2000penalty}
Munjiza, A., Andrews, K. R.~F., 2000. Penalty function method for combined
  finite--discrete element systems comprising large number of separate bodies.
  International Journal for Numerical Methods in Engineering 49~(11),
  1377--1396.

\bibitem[{Munjiza et~al.(1995)Munjiza, Owen, and Bicanic}]{munjiza1995combined}
Munjiza, A., Owen, D. R.~J., Bicanic, N., 1995. A combined finite-discrete
  element method in transient dynamics of fracturing solids. Engineering
  Computations 12~(2), 145--174.

\bibitem[{Ni{\~n}o and Garc{\'\i}a(1994)}]{nino1994gravel}
Ni{\~n}o, Y., Garc{\'\i}a, M., 1994. Gravel saltation: 2. modeling. Water
  resources research 30~(6), 1915--1924.

\bibitem[{Ni{\~n}o and Garc{\'\i}a(1998)}]{nino1998using}
Ni{\~n}o, Y., Garc{\'\i}a, M., 1998. Using lagrangian particle saltation
  observations for bedload sediment transport modelling. Hydrological Processes
  12~(8), 1197--1218.

\bibitem[{P{\"a}htz et~al.(2020)P{\"a}htz, Clark, Valyrakis, and
  Dur{\'a}n}]{pahtz2020physics}
P{\"a}htz, T., Clark, A.~H., Valyrakis, M., Dur{\'a}n, O., 2020. The physics of
  sediment transport initiation, cessation, and entrainment across aeolian and
  fluvial environments. Reviews of Geophysics 58~(1), e2019RG000679.

\bibitem[{Papanicolaou et~al.(2008)Papanicolaou, Elhakeem, Krallis, Prakash,
  and Edinger}]{papanicolaou2008sediment}
Papanicolaou, A. T.~N., Elhakeem, M., Krallis, G., Prakash, S., Edinger, J.,
  2008. Sediment transport modeling review —- current and future
  developments. Journal of Hydraulic Engineering 134~(1), 1--14.

\bibitem[{Peng et~al.(2019)Peng, Ayala, and Wang}]{peng2019direct}
Peng, C., Ayala, O.~M., Wang, L.-P., 2019. A direct numerical investigation of
  two-way interactions in a particle-laden turbulent channel flow. Journal of
  Fluid Mechanics 875, 1096--1144.

\bibitem[{Picano et~al.(2015)Picano, Breugem, and Brandt}]{picano2015turbulent}
Picano, F., Breugem, W.-P., Brandt, L., 2015. Turbulent channel flow of dense
  suspensions of neutrally buoyant spheres. Journal of Fluid Mechanics 764,
  463--487.

\bibitem[{Pozzetti et~al.(2019)Pozzetti, Jasak, Besseron, Rousset, and
  Peters}]{pozzetti2019parallel}
Pozzetti, G., Jasak, H., Besseron, X., Rousset, A., Peters, B., 2019. A
  parallel dual-grid multiscale approach to {CFD--DEM} couplings. Journal of
  Computational Physics 378, 708--722.

\bibitem[{Rana et~al.(2021)Rana, Anderson, and Day}]{ranaentrainment}
Rana, S., Anderson, W., Day, M., 2021. An entrainment paradox: how hysteretic
  saltation and secondary transport augment atmospheric uptake of aeolian
  source materials. Journal of Geophysical Research: Atmospheres 126~(10),
  e2020JD033493.

\bibitem[{Raupach et~al.(1991)Raupach, Antonia, and
  Rajagopalan}]{raupach1991rough}
Raupach, M.~R., Antonia, R.~A., Rajagopalan, S., 1991. Rough-wall turbulent
  boundary layers. Applied Mechanics Review 44~(1), 1--25.

\bibitem[{Schiller and Naumann(1933)}]{schiller1933grundlegenden}
Schiller, L., Naumann, A., 1933. {\"U}ber die grundlegenden berechnungen bei
  der schwerkraftaufbereitung. Z. Vereines Deutscher Inge. 77, 318--321.

\bibitem[{Shao(2008)}]{shao2008physics}
Shao, Y., 2008. Physics and modelling of wind erosion. Springer, Berlin.

\bibitem[{Shao and Li(1999)}]{shao1999numerical}
Shao, Y., Li, A., 1999. Numerical modelling of saltation in the atmospheric
  surface layer. Boundary-Layer Meteorology 91~(2), 199--225.

\bibitem[{Singh et~al.(2007)Singh, Sandham, and Williams}]{singh2007numerical}
Singh, K., Sandham, N., Williams, J., 2007. Numerical simulation of flow over a
  rough bed. Journal of Hydraulic Engineering 133~(4), 386--398.

\bibitem[{Tanaka et~al.(2002)Tanaka, Nishida, Kunimochi, and
  Takagi}]{tanaka2002discrete}
Tanaka, K., Nishida, M., Kunimochi, T., Takagi, T., 2002. Discrete element
  simulation and experiment for dynamic response of two-dimensional granular
  matter to the impact of a spherical projectile. Powder technology 124~(1-2),
  160--173.

\bibitem[{Tao et~al.(2018)Tao, Zhang, Guo, and Wang}]{tao2018combined}
Tao, S., Zhang, H., Guo, Z., Wang, L.-P., 2018. A combined immersed boundary
  and discrete unified gas kinetic scheme for particle--fluid flows. Journal of
  Computational Physics 375, 498--518.

\bibitem[{Tenneti and Subramaniam(2014)}]{tenneti2014particle}
Tenneti, S., Subramaniam, S., 2014. Particle-resolved direct numerical
  simulation for gas-solid flow model development. Annual Review of Fluid
  Mechanics 46, 199--230.

\bibitem[{Tschisgale et~al.(2017)Tschisgale, Kempe, and
  Fr{\"o}hlich}]{tschisgale2017non}
Tschisgale, S., Kempe, T., Fr{\"o}hlich, J., 2017. A non-iterative immersed
  boundary method for spherical particles of arbitrary density ratio. Journal
  of Computational Physics 339, 432--452.

\bibitem[{Tsuji et~al.(2008)Tsuji, Yabumoto, and Tanaka}]{tsuji2008spontaneous}
Tsuji, T., Yabumoto, K., Tanaka, T., 2008. Spontaneous structures in
  three-dimensional bubbling gas-fluidized bed by parallel {DEM--CFD} coupling
  simulation. Powder Technology 184~(2), 132--140.

\bibitem[{Uhlmann(2004)}]{uhlmann2004simulation}
Uhlmann, M., 2004. Simulation of particulate flows multi-processor machines
  with distributed memory. Tech. Rep. CIEMAT-1039, Centro de Investigaciones
  Energeticas, Medioambientales y Tecnologicas (CIEMAT), Madrid (Spain).

\bibitem[{Uhlmann(2005)}]{uhlmann2005immersed}
Uhlmann, M., 2005. An immersed boundary method with direct forcing for the
  simulation of particulate flows. Journal of Computational Physics 209~(2),
  448--476.

\bibitem[{Valance et~al.(2015)Valance, Rasmussen, El~Moctar, and
  Dupont}]{valance2015physics}
Valance, A., Rasmussen, K.~R., El~Moctar, A.~O., Dupont, P., 2015. The physics
  of {A}eolian sand transport. Comptes Rendus Physique 16~(1), 105--117.

\bibitem[{Valero-Lara(2014)}]{valero2014accelerating}
Valero-Lara, P., 2014. Accelerating solid--fluid interaction based on the
  immersed boundary method on multicore and gpu architectures. The Journal of
  Supercomputing 70~(2), 799--815.

\bibitem[{Vowinckel et~al.(2016)Vowinckel, Jain, Kempe, and
  Fr{\"o}hlich}]{vowinckel2016entrainment}
Vowinckel, B., Jain, R., Kempe, T., Fr{\"o}hlich, J., 2016. Entrainment of
  single particles in a turbulent open-channel flow: A numerical study. Journal
  of Hydraulic Research 54~(2), 158--171.

\bibitem[{Wang et~al.(2017{\natexlab{a}})Wang, Abbas, and
  Climent}]{wang2017modulation}
Wang, G., Abbas, M., Climent, {\'E}., 2017{\natexlab{a}}. Modulation of
  large-scale structures by neutrally buoyant and inertial finite-size
  particles in turbulent {C}ouette flow. Physical Review Fluids 2~(8), 084302.

\bibitem[{Wang et~al.(2013)Wang, He, and Zhang}]{wang2013parallel}
Wang, S., He, G., Zhang, X., 2013. Parallel computing strategy for a flow
  solver based on immersed boundary method and discrete stream-function
  formulation. Computers \& Fluids 88, 210--224.

\bibitem[{Wang et~al.(2017{\natexlab{b}})Wang, Luo, Yang, Hu, and
  Fan}]{wang2017parallel}
Wang, S., Luo, K., Yang, S., Hu, C., Fan, J., 2017{\natexlab{b}}. Parallel
  {LES--DEM} simulation of dense flows in fluidized beds. Applied Thermal
  Engineering 111, 1523--1535.

\bibitem[{Wang et~al.(2019)Wang, Vanella, and Balaras}]{wang2019hydrodynamic}
Wang, S., Vanella, M., Balaras, E., 2019. A hydrodynamic stress model for
  simulating turbulence/particle interactions with immersed boundary methods.
  Journal of Computational Physics 382, 240--263.

\bibitem[{Werner(1990)}]{werner1990steady}
Werner, B., 1990. A steady-state model of wind-blown sand transport. The
  Journal of Geology 98~(1), 1--17.

\bibitem[{Wiberg and Dungan~Smith(1989)}]{wiberg1989model}
Wiberg, P.~L., Dungan~Smith, J., 1989. Model for calculating bed load transport
  of sediment. Journal of Hydraulic Engineering 115~(1), 101--123.

\bibitem[{Wiberg and Smith(1985)}]{wiberg1985theoretical}
Wiberg, P.~L., Smith, J.~D., 1985. A theoretical model for saltating grains in
  water. Journal of Geophysical Research: Oceans 90~(C4), 7341--7354.

\bibitem[{Yang and Balachandar(2021)}]{yang2021scalable}
Yang, Y., Balachandar, S., 2021. A scalable parallel algorithm for
  direct-forcing immersed boundary method for multiphase flow simulation on
  spectral elements. The Journal of Supercomputing 77~(3), 2897--2927.

\bibitem[{Yang et~al.(2018)Yang, Deng, and Shen}]{yang2018direct}
Yang, Z., Deng, B.-Q., Shen, L., 2018. Direct numerical simulation of wind
  turbulence over breaking waves. Journal of Fluid Mechanics 850, 120--155.

\bibitem[{Yang et~al.(2017)Yang, Lu, Guo, Liu, and Shen}]{yang2017}
Yang, Z., Lu, X.-H., Guo, X., Liu, Y., Shen, L., 2017. Numerical simulation of
  sediment suspension and transport under plunging breaking waves. Computers \&
  Fluids 158, 57--71.

\bibitem[{Yu and Shao(2007)}]{yu2007direct}
Yu, Z., Shao, X., 2007. A direct-forcing fictitious domain method for
  particulate flows. Journal of Computational Physics 227~(1), 292--314.

\bibitem[{Yu et~al.(2006)Yu, Shao, and Wachs}]{yu2006fictitious}
Yu, Z., Shao, X., Wachs, A., 2006. A fictitious domain method for particulate
  flows with heat transfer. J. Comput. Phys. 217~(2), 424--452.

\bibitem[{Zhang et~al.(2018)Zhang, Hu, and Zheng}]{zhang2018large}
Zhang, Y., Hu, R., Zheng, X., 2018. Large-scale coherent structures of
  suspended dust concentration in the neutral atmospheric surface layer: A
  large-eddy simulation study. Physics of Fluids 30~(4), 046601.

\bibitem[{Zheng(2009)}]{zheng2009mechanics}
Zheng, X., 2009. Mechanics of wind-blown sand movements. Springer, Berlin.

\bibitem[{Zheng et~al.(2021{\natexlab{a}})Zheng, Feng, and
  Wang}]{zheng2021modulation}
Zheng, X., Feng, S., Wang, P., 2021{\natexlab{a}}. Modulation of turbulence by
  saltating particles on erodible bed surface. Journal of Fluid Mechanics 918,
  A16.

\bibitem[{Zheng et~al.(2021{\natexlab{b}})Zheng, Wang, and
  Zhu}]{zheng2021experimental}
Zheng, X., Wang, G., Zhu, W., 2021{\natexlab{b}}. Experimental study on the
  effects of particle--wall interactions on {VLSM} in sand-laden flows. Journal
  of Fluid Mechanics 914, A35.

\bibitem[{Zhou and Fan(2014)}]{zhou2014second}
Zhou, Q., Fan, L.-S., 2014. A second-order accurate immersed boundary-lattice
  {B}oltzmann method for particle-laden flows. Journal of Computational Physics
  268, 269--301.

\bibitem[{Zhu et~al.(2021)Zhu, Pan, Wang, Liang, Ji, and Wang}]{zhu2021AE}
Zhu, H., Pan, C., Wang, G., Liang, Y., Ji, X., Wang, J., 2021. Attached
  eddy-like particle clustering in a turbulent boundary layer under net
  sedimentation conditions. Journal of Fluid Mechanics 920, A53.

\bibitem[{Zhu et~al.(2019)Zhu, Pan, Wang, Liang, and Ji}]{zhu2019sand}
Zhu, H.-Y., Pan, C., Wang, J.-J., Liang, Y.-R., Ji, X.-C., 2019.
  Sand-turbulence interaction in a high-{R}eynolds-number turbulent boundary
  layer under net sedimentation conditions. International Journal of Multiphase
  Flow 119, 56--71.

\end{thebibliography}






\end{document}